\documentclass{article}

\usepackage{enumerate}
\usepackage{amsfonts}
\usepackage{amsmath}
\usepackage{comment}
\usepackage{textcomp}
\usepackage{amsthm, tikz}
\usepackage{amssymb}
\usepackage{color}
\usepackage{graphicx}
\usepackage{bbm}
\usepackage[matrix,arrow,curve]{xy}
\usepackage{array}
\usepackage{mathtools}

\def\be{\begin{eqnarray}}
\def\ee{\end{eqnarray}}
\def\nn{\nonumber}

\def\Tr{{\rm Tr}\,}
\def\Z{\mathbb{Z}}

\def\l[{\phantom.[}
\def\lm{\limits}

\def\0{\emptyset}

\hoffset= - 1.0in         

\begin{document}

\title{{\bf {Knot invariants from Virasoro related representation and pretzel knots
}\vspace{.2cm}}
\author{{\bf D. Galakhov$^{a,b,}$}\footnote{galakhov@itep.ru; galakhov@physics.rutgers.edu}, \ {\bf D. Melnikov$^{a,c,}$}\footnote{dmitry@iip.ufrn.br}, \  {\bf A. Mironov$^{d,a,e,}$}\footnote{mironov@lpi.ru; mironov@itep.ru}, \ and \ {\bf A. Morozov$^{a,e,}$}\thanks{morozov@itep.ru}}
\date{ }
}

\maketitle

\vspace{-5.5cm}

\begin{center}
\hfill FIAN/TD-21/14\\
\hfill ITEP/TH-02/15\\
\end{center}

\vspace{4.2cm}

\begin{center}
$^a$ {\small {\it ITEP, Moscow 117218, Russia}}\\
$^b$ {\small {\it NHETC and Department of Physics and Astronomy, Rutgers University,
Piscataway, NJ 08855-0849, USA }}\\
$^c$ {\small {\it International Institute of Physics, UFRN
Av. Odilon G. de Lima 1722, Natal 59078-400, Brazil}}\\
$^d$ {\small {\it Lebedev Physics Institute, Moscow 119991, Russia}}\\
$^e$ {\small {\it National Research Nuclear University MEPhI, Moscow 115409, Russia }}
\end{center}

\vspace{1cm}

\begin{abstract}
We remind the method to calculate colored Jones polynomials
for the plat representations of knot diagrams from the knowledge
of modular transformation (monodromies) of
Virasoro conformal blocks with insertions of degenerate fields.
As an illustration we use a rich family of pretzel knots,
lying on a surface of arbitrary genus $g$,
which was recently analyzed by the evolution method.
Further generalizations can be to generic Virasoro
modular transformations, provided by integral kernels,
which can lead to the Hikami invariants.
\end{abstract}

\vspace{2cm}

\section{Introduction}

Knot polynomials are Wilson loop averages in $3d$ Chern-Simons theory \cite{CS,Wit},
\be
{\cal H}^{\cal L}_R(q,G)\ = \ \left< \
\Tr_R P\exp\left(\oint_{\cal L} {\cal A}\right) \ \
\ \exp\left\{\frac{2\pi i}{\hbar} \int_{\cal M} \Tr_{Adj}
\Big({\cal A}d{\cal A}+\frac{2}{3}{\cal A}^3\Big)\right\}  \
\right>
\ee
where ${\cal A}$ is the $G$-connection on a $3d$ manifold ${\cal M}$, to which a line
(knot or link) ${\cal L}$ belongs.
They depend only on topology of ${\cal L}$, and also on the coupling constant
$q=e^\hbar$ and representation $R$ of the Lie algebra $G$.
According to \cite{Wit} knot polynomials can be interpreted in terms of monodromies
of conformal blocks in $2d$ conformal theory on the equal-time slice $\Sigma$ of ${\cal M}$.
In the simplest case of ''ordinary'' knots and links ${\cal M} \approx R^3$
and often also $\Sigma = S^2$.
A possible way to less-trivial ${\cal M}$ is via consideration of
``virtual'' knots, {\it a la} \cite{Kaufvirt}.

Today there are two effective ways to calculate ``ordinary'' knot polynomials:
by the Reshetikhin-Turaev (RT) group theory method \cite{RT,inds,MMMkn2,MMMknots},
using explicit
quantum ${\cal R}$-matrices in various representations of various groups
(well-known skein relations are a simple particular case of this),
and by Khovanov's hypercube method \cite{Kho} in the modified version of
\cite{DM3}, applicable to algebra $G=SU(N)$ with arbitrary $N$.
Both approaches are partly related, first, via peculiar Kauffman's
${\cal R}$-matrix for $N=2$ \cite{KaufR} and, more generally, in
\cite{AnoMDM3}.
However, Khovanov's method is currently more universal:
applicable also  to virtual knots \cite{MMMvirt}
and to additional $\beta$-deformation \cite{betadefo} of HOMFLY
to super-polynomials \cite{suppol,DMMSS}.
Instead the RT approach is much better suited for the study of
representation dependence of knot polynomials and, after being combined
with the powerful evolution method \cite{evo}, it allows one to find
explicit formulas for various families of knots and links.

In this paper we provide one more important illustration.
Namely, instead of the ordinary quantum Racah matrices for $SU_q(N)$ algebras,
associated with monodromies of conformal blocks in the WZNW theory \cite{WZNW,AG},
which were widely used in \cite{inds}, we follow \cite{GMMlast} and take the
${\cal R}$-matrices induced by the modular transformations $S$ and $T$
for the Virasoro (Liouville) conformal blocks \cite{CFT}.
This provides an additional powerful method to calculate the Jones polynomials,
which is important, first of all, conceptually: going from the degenerate conformal blocks to generic one,
i.e. from finite-matrix to integral modular transforms of \cite{Tesh},
i.e. switching to infinite representations (of $SL(2,C)$, \cite{GT}),
one opens a way to describe the Hikami knot invariants \cite{Hik,GMMlast}
and this could be a first step towards constructing knot polynomials
for infinite representations.
As to practice, we apply the method to calculate Jones polynomials
for a rich family of knots that can be put on genus-$g$ surface and are called pretzel knots, recently introduced and studied
by the evolution method in \cite{Sle}.

Further extension of the conformal field theory technique to $N>2$ involves understanding of modular transformation
formulas for $W$-algebras, what is a work in progress \cite{Fabio}.
This is a particularly interesting question also because the
pretzel knot family contains a vast set of mutant knots,
which are undistinguishable at the level of (anti)symmetric representations,
and, in particular, by any colored Jones polynomials.

\section{The basic ideas}
\label{sec:STmatrices}

\subsection{Knots and $PSL(2,\mathbb{Z})$}
\label{sec:ST}

It is well known that knots can be presented as trace-like closures of braids. Knot invariants in this case can be constructed as traces of the corresponding braid group elements:
\be
\mbox{Inv}(b)=\Tr_{\!R}\, \beta(b)\,, \qquad b\in B_n\,,\qquad  R:b\to \beta\,
\ee
Here $b$ is an element of Artin's braid group of $n$ elements $B_n$ and the trace is taken in some representation $R$. For this to give an invariant the trace operation must respect the two Markov moves. The first Markov move ensures that the operation $\Tr$ is indeed a closure of the braid, i.e. it is a trace, $\Tr gbg^{-1}=\Tr b$. The second Markov move guarantees that $\Tr$ yields an invariant. Specifically, it is equivalent to the first Reidemeister move, while the other two Reidemeister moves are automatically satisfied in the braid group.

It is also known that the modular group $PSL(2,\mathbb{Z})$ yields a representation of the braid group on 3 elements $B_3$. Indeed $B_3$ has two generators $b_1$ and $b_2$ satisfying the Yang-Baxter relation
\be
b_1b_2b_1=b_2b_1b_2\,
\ee
In turn $PSL(2,\Z)$ can be generated by elements $S$ and $T$ satisfying two relations
\be
\label{STcube}
S^2=1\,, \qquad (ST)^3=1\,
\ee
Choosing
\be
\label{B3rep}
\beta(b_1)=T\,, \qquad \beta(b_2)=STS\,,
\ee
one constructs a representation of $B_3$. Notice however that $(b_1b_2)^3$ is the central element of $B_3$, while in the given representation one has the relation $\beta((b_1b_2)^3)=(ST)^6=1$. The module generated by~(\ref{STcube}) is not free and~(\ref{B3rep}) is not the most general solution of the Yang-Baxter relation.

In what follows the ansatz~(\ref{B3rep}) will correspond to the simplest situation of the Virasoro four-point spherical conformal block. More generally, if one thinks of $S$ as a unitary transformation acting on the operator $T$,
\be
\label{B3rep2}
\beta(b_1) = T\,, \qquad   \beta(b_2) = STS^\dagger\,,
\ee
which is consistent with the Yang-Baxter relation provided that
\begin{eqnarray}
SS^\dagger & = & 1, \nn \\
STS^\dagger TST & = & 1\,.
\label{STrels0}
\end{eqnarray}
This representation is subject to the constraint $\beta((b_1b_2)^3)=1$. For $SU(N)$ the second formula is no longer correct, and one
needs two $T$- and $S$-matrices so that the second formula generalizes to (see section 2.2)
\be\label{mod}
S\bar T \bar S\bar T S^\dagger T S = TS\bar T\bar S\bar T
\ee

The Yang-Baxter relation imposes that the third Reidemeister move is respected. The second Reidemeister move is trivially satisfied for the representation~(\ref{B3rep2}) as this is simply the statement $b_1^{-1}b_1=b_2^{-1}b_2=1$.
The consistency with the first Reidemeister move should follow from the definition of the trace, applied to some matrix-valued $S$ and $T$. The appropriate trace is provided by the RT formalism~\cite{MMMkn2}, e.g. selecting
\be
\label{rep}
\boxed{
R\otimes I = T , \ \ \ \ \ \ \   I\otimes R = STS^\dagger
}
\ee
where $I$ is a unity operator acting on a vector space $V$, while $R$ is acting on $V\otimes V$, and one relates $T$ with the ${\cal R}$-matrix and $S$ with the mixing matrix of the RT construction. Then, the invariance with respect to the first Reidemeister move follows from the following property of the ${\cal R}$-matrix:
\be
\label{projector}
R^{ij}_{kl} P^k_i = ({\rm tr}\,I)\,I^j_l\,,
\ee
where $P$ is the projector closing a strand of the braid.

One can generalize the above construction to the case of $B_n$ by considering $PSL(2,\mathbb{Z})$ representations of the $B_3$ subgroups generated by any pair of consecutive generators $b_i$, $b_{i+1}$. In the ${\cal R}$-matrix formalism one assigns
\begin{eqnarray}
\beta(b_1) & = & R\otimes I\otimes I\otimes I\otimes \cdots\,, \nn \\
\beta(b_2) & = & I\otimes R\otimes I\otimes I\otimes \cdots\,, \nn \\
\beta(b_3) & = & I\otimes I\otimes R\otimes I\otimes \cdots\,, \quad \text{etc}
\label{Bn}
\end{eqnarray}
In the following we will use representation~(\ref{rep}) for any two consecutive lines in~(\ref{Bn}). The $S$ and $T$ matrices corresponding to different $PSL(2,\mathbb{Z})$ will act on different vector spaces and thus their product will not be just a matrix, but a multi-matrix product.

The described version of the RT construction assumes that the braid is closed in a trace like manner. However, one can generalize it to an arbitrary closure without changing the RT definition of the trace. In fact, different closures can be related by insertions of a combination of $S$-operators of the form
\be
I\otimes I\otimes \cdots \otimes S\otimes  \cdots \otimes I,
\ee
where $S$ acts on $V\otimes V\otimes V$. To see how that works, it is useful to recall the representation of $PSL(2,\mathbb{Z})$ as a group of modular transformations as we do in the next subsection. Here one needs to show that the $S$-operators do not alter the invariance of the trace under the Reidemeister moves. Since $S$ is a unitary transformation, what it does, it changes a basis in $V\otimes V\otimes V$. The trace does not depend on the choice of the basis and thus the invariance is preserved.

From the point of view of knots, the vector spaces $V$ are associated to the knot or link lines and are not obligatory the same. They form representation spaces of the group $G$. The representations of $S$ and $T$ operators depend on those. The simplest non-trivial example of (\ref{rep}) can be constructed from the known results on ${\mathcal{R}}$-matrices and is given by $2\times 2$ matrices acting in the two-dimensional space of intertwining operators ${\cal M}$ acting from the tensor cube of the fundamental representation to the mixed representation: ${\cal M}:\ \hbox{fund}^{\otimes 3}\to\ \hbox{mixed}$. If $T$ is diagonal, then $S$ coincides with the elementary mixing matrix of \cite{MMMkn2}:
\be
T = \left(\begin{array}{cc}  -q & 0 \\ \\ 0 & \frac{1}{q}\end{array}\right), \ \ \ \ \ \
S = \left(\begin{array}{cc}  \frac{1}{[2]} & \frac{\sqrt{[3]}}{[2]} \\ \\
\frac{\sqrt{[3]}}{[2]} & -\frac{1}{[2]}\end{array}\right)
\label{ST2}
\ee
where the square brackets stand for quantum numbers: $[2]=q+q^{-1}$ and $[3] = q^2+1+q^{-2}$. For $q=1$ the matrix $S$ becomes
\be
S = \left(\begin{array}{cc} \frac{1}{2} & \frac{\sqrt{3}}{2} \\ \\
\frac{\sqrt{3}}{2} & -\frac{1}{2} \end{array} \right) =
\left(\begin{array}{cc} \cos\frac{\pi}{3} & \sin\frac{\pi}{3} \\ \\
\sin\frac{\pi}{3} & -\cos\frac{\pi}{3} \end{array} \right)
\ee
that is an orthogonal matrix of a rotation by 60 degrees. The matrix $T$ in this limit represents a reflection. The triple rotation by 60 degrees combined with the reflections gives an identity operation $(ST)^3=1$.

Since the example~(\ref{ST2}) corresponds to the case of $V$ being the fundamental representations of $SU(2)$, these two matrices provide a full description of the (uncolored) Jones polynomials, see section~\ref{sec:examples} below. To go beyond the fundamental representation of $SU(2)$, one needs a generalization of~(\ref{ST2}). There are several ways to find it. The most popular is to treat $S$ as a Racah "matrix" ($6j$-symbols) in the representation theory of $SU_q(N)$ group: this leads to a description of the colored HOMFLY polynomials, developed pretty far mostly in \cite{inds}.

From the point of view of conformal theory, this story is about modular transformations in the WZNW model \cite{AG}. We briefly remind this story in subsection~(\ref{sec:Racah}). However, from the point of view of modular transformations, the even simpler case would be those of Liouville and $W_N$-models. Thus, we begin from them in the next subsection~\ref{sec:confblocks}. The finite-size matrices $S$ and $T$ in these cases arise as monodromies of solutions to the BPZ equations, which can be derived for the conformal blocks with some fields degenerate. Monodromies of generic conformal blocks should provide, in exactly the same way, the integral Hikami invariants of knots (associated with non-finite-dimensional representations of the underlying algebra), and this makes the consideration of Virasoro/$W$-, and not just the Kac-Moody, conformal blocks an additionally interesting subject. In the finite dimensional case, however, the $SU_q(2)$ and Virasoro description are known to coincide.

\subsection{$S$ and $T$ as monodromies of BPZ equations. $(1,2)$-degenerate fields vs fundamental representation}
\label{sec:confblocks}

To demonstrate how specific representations of $S$ and $T$ operators, like the one in equation~(\ref{ST2}), can be constructed let us discuss an example of vector spaces $V$ provided by CFT's \cite{CFT}. Recall that the $PSL(2,\Z)$ group is the modular group and thus a symmetry of the CFT's. In particular it acts on the conformal blocks of a CFT.

Rational CFT's possess a finite number of primary fields as well as null operators that lead to differential equations for the correlation functions. Null operators annihilate some ``null'' states of a CFT, also called degenerate fields. In general modular transformations of conformal blocks are given by integral kernels, but for the ones containing a degenerate field the kernels become finite dimensional, i.e. finite matrices.

Consider a four-point correlation function that contains a degenerate field,
\be
\label{confblock}
\langle V(z)V_0(0)V_{1}(1)V_\infty(\infty)\rangle = F_{\alpha}(z,0,1,\infty)\,,
\ee
where $\alpha=\{\alpha_i\}\equiv\{\alpha_z,\alpha_0,\alpha_1,\alpha_\infty\}$ denote the highest weight vectors of the corresponding primaries $V_i$. The highest weights are classified by a pair of integer numbers $\alpha_i\equiv \alpha^{(i)}_{(m,n)}$, where
\be
\alpha_{(m,n)}=\frac12\left(\frac{m-1}{b}-(n-1)b\right),
\ee
while
\be
\Delta_{m,n}=\alpha_{(m,n)}\left(\alpha_{(m,n)}-b+\frac1b\right), \qquad \text{and} \qquad c=1-6\left(b-\frac{1}{b}\right)^2
\ee
are the conformal weight of the field and central charge of the CFT respectively.

If $V$ is a degenerate field (at the level $m\cdot n$), which is annihilated by a null operator of the order $m\cdot n$, one can use the Ward identities to derive a differential equations for the 4-point function. For example, in the simplest case of $(m,n)=(1,2)$ one has the level-two null operator $L_2+{b^2}\,L_{-1}^2$, which leads to the BPZ equation \cite{CFT}
\be
(L_2+{b^2}\,L_{-1}^2)F_\alpha(z)=0\,
\ee
and consequently, to a hypergeometric equation
\be
\left(z(1-z)\frac{d^2}{dz^2}+\left(C-(A+B+1)z\right)\frac{d}{dz}-AB\right)F(z)=0\,
\ee
after the function redefinition $F_\alpha(z)=x^\delta(1-x)^{\bar\delta}F(z)$. See~\cite{CFT,Fateev,surop,GMMlast} for the details of the Virasoro and $W_N$ cases.

The differential equation is invariant under the modular transformation, but not necessarily its solutions. In the case of the second order differential equation above, the general solution is given by linear combinations of two independent solutions, which are transformed one into another under the action of the modular group. Natural linear combinations (bases) of the solutions are provided by the functions that have analytic behavior if some of the points come close together. In particular, for $z\to 0$ and $z\to 1$ in the equation above one has two sets of solutions that have good Laurent expansions:
\be
\label{solns}
\left\{
\begin{array}{c}
 _2F_1(A,B;C|z)\\
 _2F_1(A-C+1,B-C+1;2-C|z)
\end{array}
\right\}\, \qquad \text{or} \qquad
\left\{
\begin{array}{c}
 _2F_1(A,B;A+B-C+1|1-z)\\
 _2F_1(C-A,C-B;C-A-B+1|1-z)
\end{array}
\right\}\,
\ee

Since the space $V$ is here two-dimensional, we associate this case of $(1,2)$ degenerate fields with the case of fundamental representations of $SU(2)$. Similarly, in the case of $(1,k)$ degenerate fields the space of solutions is $k$-dimensional and is associated with higher representations of $SU(2)$.

The $T$-transformation corresponds to an exchange of two points, say $z$ and $0$, that is it generates a (half-) monodromy. While the monodromy is diagonal in one basis, it is a non-diagonal matrix permuting the two solutions in the second basis. In general, we will denote $T$ the diagonal half-monodromy matrix
\be
\label{Tgen}
T(\alpha_1,\alpha_2)=\left(
\begin{array}{cccc}
 \lambda_0 & 0 &  & 0 \\
 0 & \lambda_1 &  & 0 \\
  &  & \ddots & \vdots \\
 0 & 0 & \cdots & \lambda_{n} \\
\end{array}
\right)\,
\ee
where the dependence of $T$ on the representations of the external fields is shown explicitly. The eigenvalues of $T$ are known:
\be\label{fr}
\lambda_i(R_1,R_2)=\pm\, q^{C_{R_i}-C_{R_1}-C_{R_2}}\,
\ee
where $C_R$ is the quadratic Casimir in the representation $R$. In particular, for the two fundamental reps one derives the $T$-matrix from equation~(\ref{ST2}) up to a normalization and signs.

The $S$-matrix can be then found as an appropriate unitary transformation that relates the two bases and diagonalizes $T$ for the second basis:
\be
B_{R_s}\left[\begin{array}{cc}
    R_0 & R_1\\
    R_z & R_\infty\\
\end{array}\right](z)=\sum\limits_{R_t} S_{R_s R_t}\left[\begin{array}{cc}
R_0 & R_1\\
R_z & R_\infty\\
\end{array}\right]B_{R_t}\left[\begin{array}{cc}
R_0 & R_1\\
R_z & R_\infty\\
\end{array}\right](1-z)\,
\ee
where the notations $B_{R_i}[\cdot](x)$ for the eigenbases of the conformal blocks near zero $x=z$ and one $x=1-z$ generalize the sets of solutions in equation~(\ref{solns}).

This has the following diagrammatic presentation
{\unitlength 1mm 
    \linethickness{0.4pt}
    \be
    \begin{picture}(40,15)(0,15)
    \put(5,5){\line(1,1){10}}
    \put(5,25){\line(1,-1){10}}
    \put(15,15){\line(1,0){10}}
    \put(25,15){\line(1,1){10}}
    \put(25,15){\line(1,-1){10}}
    \put(12,9){\mbox{$V_z$}}
    \put(12,19){\mbox{$V_0$}}
    \put(18,16){\mbox{$V_s$}}
    \put(24,9){\mbox{$V_\infty$}}
    \put(24,19){\mbox{$V_1$}}
    \end{picture}\quad =\quad \sum\lm_{R_t} \ \ S_{R_s R_t}\left[\begin{array}{cc}
R_0 & R_1\\
R_z & R_\infty\\
\end{array}\right]\quad
    \begin{picture}(30,20)(0,20)
    \put(5,5){\line(1,1){10}}
    \put(5,35){\line(1,-1){10}}
    \put(15,15){\line(0,1){10}}
    \put(15,25){\line(1,1){10}}
    \put(15,15){\line(1,-1){10}}
    \put(7,12){\mbox{$V_z$}}
    \put(19,12){\mbox{$V_\infty$}}
    \put(16,19){\mbox{$V_t$}}
    \put(7,26){\mbox{$V_0$}}
    \put(19,26){\mbox{$V_1$}}
    \end{picture}
    \ee
    \vspace{15mm}
}

In the case of fundamental reps of $SU(2)$ one gets
\be\label{Sf}
S=\left(
\begin{array}{cc}
    \frac{\Gamma \left(2+2/b^2\right) \Gamma \left(-1-2/b^2\right)}{\Gamma \left(1+1/b^2\right) \Gamma \left(-1/b^2\right)} & \frac{\Gamma \left(2+2/b^2\right) \Gamma \left(1+2/b^2\right)}{\Gamma \left(1+1/b^2\right) \Gamma \left(2+3/b^2\right)} \\
    & \\
    \frac{\Gamma \left(-2/{b^2}\right) \Gamma \left(-1-2/{b^2}\right)}{\Gamma \left(-1/{ b^2}\right) \Gamma \left(-1-3/b^2\right)} & \frac{\Gamma \left(-2/{b^2}\right) \Gamma \left(1+2/{b^2}\right)}{\Gamma \left(-1/{b^2}\right) \Gamma \left(1+1/{b^2}\right)}
\end{array}
\right)
= U\left(\begin{array}{cc}  \frac{1}{[2]} & \frac{\sqrt{[3]}}{[2]} \\ \\
\frac{\sqrt{[3]}}{[2]} & -\frac{1}{[2]}\end{array}
\right)U^{-1}\,
\ee
which is the same $S$-matrix as in~(\ref{ST2}) up to a unitary rotation $U$ \cite{GMMlast}. Here we used the relation $q=e^{i\pi/b^2}$.

In the above discussion the four-point conformal blocks, as solutions of a differential equation, have furnished a representation of the $S$ and $T$ operations. Independent solutions of the equation spanned the vector space $V\otimes V\otimes V$ of the previous section. Action of the $T$ matrix, interchanging the insertion points of the primary fields is equivalent to the braiding operation. The basis of conformal blocks can be made orthonormal, so that the trace operation, necessary to construct a knot invariant, can be realized through the projectors like~(\ref{projector}) made out of the normalized conformal blocks.

In the general case, one needs to consider conformal blocks with an arbitrary number of insertions points. One can use one set of $S$ and $T$ matrices for the given representations of the fields inserted in any three close points in order to perform arbitrary braidings. Here close mean that we presumably work in a basis of conformal blocks labeled by the tensor product $V\otimes V\otimes V$. The braidings will appear as some words of $S$ and $T$ matrices. For another triple of close points one will use another set of matrices, so that the total operation on the conformal blocks will be a tensor, or a multi-matrix products of words of $S$ and $T$, applied in different channels. We will demonstrate this method in a number of examples in sections~\ref{sec:examples} and~\ref{sec:generalform}.

Let us stress here again that the answer (\ref{Sf}) obtained in CFT coincides with the $SU_q(2)$ formula for the fundamental representations (\ref{ST2}) only up to a matrix $U$ \cite[eqs.(5.27)-(5.28)]{GMMlast}. Such a matrix $S$, however, still satisfies the basic relations (\ref{STcube}) and is due to chosen normalization of the conformal block such that the small-$z$ expansion of the conformal block starts from unity. It can be removed by a properly chosen normalization presumably consistent with \cite{Tesh}.

In the presented method of computing the knot invariants, one only uses the fact that the conformal blocks are normalized. The only obstruction in the computations are the $S$ and $T$ matrices themselves, which follow from the modular properties of the conformal blocks. In general deriving the differential equation for the 4-point function is not straightforward. For the $N=2$ (Virasoro) case, to have enough constraints to derive the BPZ equation it is enough to assume that one of the fields is degenerate, for the general $W_N$ with $N>2$ however, one has to make additional assumptions about other fields as well. The order of the differential equation will grow with $N$ as well as with dimension of the degenerate field.

In some particular cases simplifications are possible. For example, if all the fields of the conformal block are in the fundamental, or antifundamental representations the BPZ equation can be reduced to a second order hypergeometric equation. This fact was used in ref.~\cite{Fabio} to derive the $S$ and $T$ matrices for general $N$. Specifically, for the choice $V$, $V_0$ are both fundamentals and $V_1$ is anti-fundamental, one derives
\be
S=\frac{1}{\sqrt{[2][N]}}\,U\left(
\begin{array}{cc}
\sqrt{[N-1]} & \sqrt{[N+1]} \\
\sqrt{[N+1]} & - \sqrt{[N-1]}
\end{array}
\right)U^{-1}\,
\ee
while for $V$ fundamental and $V_0$ and $V_1$ anti-fundamental, one gets
\be
\bar S=\frac{1}{[N]}\,U\left(
\begin{array}{cc}
1 & \sqrt{[N-1][N+1]} \\
\sqrt{[N-1][N+1]} & - 1
\end{array}
\right)U^{-1}\,
\ee
Here again the rotation matrix $U$ can be removed by a proper normalization of conformal blocks \cite{Fabio}, and then these matrices satisfy (\ref{mod}) with
\be
T=\left(
\begin{array}{cc}
q^{1-N} & 0 \\
0 & - q^{-1-N}
\end{array}
\right)
\ \ \ \ \ \ \ \ \ \ \
\bar T=\left(
\begin{array}{cc}
1 & 0 \\
0 & - q^N
\end{array}
\right)
\ee

For higher representations, the $S$ matrices can be derived from more basic ones, using the general properties of the $S$ matrices from the group-theoretical point of view. We will provide some examples in the next subsection.

\subsection{ $S$ as a Racah matrix. Higher representations}
\label{sec:Racah}

As we already mentioned, one can discuss the same $S$-matrices from the point of view of representation theory~\cite{MooreSeiberg}. The primary fields are classified by the irreps and the product of the vector spaces $V$ is the tensor product of representations. The independent solutions of the BPZ differential equations are labelled by the irreps that appear in the tensor product $V\otimes V$.

There is a natural isomorphism coming from the associativity of the tensor product. More specifically, for three representations $R_i$, $R_j$ and $R_k$ one can derive
\begin{eqnarray}
R_i\otimes R_j \otimes R_k & \longrightarrow & \sum\limits_n N_{ij}^n \, R_n\otimes R_k \longrightarrow \sum\limits_{l,n}N_{ij}^nN_{nk}^l \,R_l\\
&\longrightarrow& \sum\limits_m N_{jk}^m R_i\otimes R_m \longrightarrow \sum\limits_{l,m}N_{jk}^mN_{im}^l \,R_l\,
\end{eqnarray}
where the arrows denote the action of intertwining operators and the sums over .
According to the order, in which the tensor product is taken, the intermediate sum in both lines goes over different (in general) sets of irreps $R_m$ and $R_n$. Notice that the coefficients $N_{ij}^k$, which appear here are known on the CFT side as the fusion coefficients~\cite{Verlinde,MooreSeiberg}.

The isomorphism is a transformation relating the corresponding vector spaces
\be
\label{isomorphism}
S_{mn}\left[
\begin{array}{cc}
R_j & R_k \\
R_i & R_l
\end{array}
\right]: \sum\limits_{n}N_{ij}^nN_{nk}^l \to \sum\limits_{m}N_{jk}^mN_{im}^l\,
\ee
where the matrix $S$ is also called Racah or duality matrix. It is also related to Wigner's $6j$ symbols. By definition, $S$ is a unitary transformation and $S^\dagger S=1$, and the Racah matrices form a representation of the operators $S$ of section~\ref{sec:ST}.

In what follows we consider the representations of $SU_q(N)$ with $|q|\ne 1$ and label them by partitions $[\lambda_1\lambda_2\ldots \lambda_{N-1}]$, or equivalently by the Young diagrams. In the case of $SU_q(2)$ the partition is characterized by only one number, which is equivalent to the doubled spin of the representation. We will sometimes omit the square brackets and, somewhat misleadingly, identify $\lambda$ and the spin $j$, so that the fundamental representation corresponds to $j=1$. As a simple illustration of~(\ref{isomorphism}) consider the representation-product diagrams from~\cite{MMMkn2} for the particular choice of external legs:

\begin{picture}(300,100)(-150,-30)
\put(0,0){\line(0,-1){20}}
\put(0,0){\line(-1,1){40}}
\put(0,0){\line(1,1){40}}
\put(-20,20){\line(1,1){20}}
\put(200,0){\line(0,-1){20}}
\put(200,0){\line(-1,1){40}}
\put(200,0){\line(1,1){40}}
\put(220,20){\line(-1,1){20}}
\put(5,-22){\mbox{$[1]$}}
\put(205,-22){\mbox{$[1]$}}
\put(-46,46){\mbox{$[1]$}}
\put(-4,46){\mbox{$[1]$}}
\put(38,46){\mbox{$\overline{[1]}$}}
\put(154,46){\mbox{$[1]$}}
\put(196,46){\mbox{$[1]$}}
\put(238,46){\mbox{$\overline{[1]}$}}
\put(-15,2){\mbox{$i$}}
\put(212,2){\mbox{$J$}}
\put(40,13){\mbox{$= \ \ \ \sum_J \  S_{iJ}\left[
\begin{array}{cc} [1] & \overline{[1]} \\ {[1]} & \overline{[1]}\end{array}
\right]$}}
\end{picture}

\noindent
Both sets of intermediate states are $2$-dimensional, but different: $i=[2],[11]$ and $J=0,Adj=[0],[21^{N-1}]$. Note also that the conjugate fundamental representation $\overline{[1]} = [1^{N-1}]$. For $N=2$ however the two sets coincide: $\overline{[1]}=[1]$, $Adj=[2]$, $[11]=[0]$, therefore the two diagrams are the same. The matrix $S_{iJ}$ in this case coincides with the Racah matrix for $[1]^{\otimes 3}\to [1]$, which is known from ref.~\cite{MMMkn2} to be exactly (\ref{ST2}).

As we will encounter in the following sections, the Racah matrices have some obvious symmetry properties under the permutation of the representations and the Biedenharn-Elliot sum rule, which is also known as the pentagon identity~\cite{MooreSeiberg}:
\be
\label{pentagon}
S_{mn}\left[
\begin{array}{cc}
R_j & R_k \\
R_i & R_l
\end{array}
\right]=\sum\limits_{r_1,r_2} S_{jr_1}\left[
\begin{array}{cc}
R_b & R_m \\
R_a & R_i
\end{array}
\right]S_{mr_2}\left[
\begin{array}{cc}
R_b & R_k \\
R_{r_1} & R_{l}
\end{array}
\right]S_{r_1n}\left[
\begin{array}{cc}
R_a & R_{r_2} \\
R_i & R_{l}
\end{array}
\right]S_{r_2j}\left[
\begin{array}{cc}
R_a & R_b \\
R_n & R_k
\end{array}
\right]
\ee
These properties allow one to compute the unknown Racah matrices explicitly in some simple cases. Let us demonstrate this in the simple example of $SU_q(2)$.  In fact, what we are going to show is that some of the Racah matrices can always be reduced to the ones containing one fundamental representation:
\be
\label{RacahwFund}
S_{st}\left(\begin{array}{cc} [1] & R_3 \\ R_1 & R_4\end{array}\right)
\ee
We will consider the case of $SU(2)$, for which $[1]\equiv 1=j_2$, $R_1\equiv j_1$ and so on. In the latter case, the $S$-matrices with arbitrary integer $j_2$ can be reduced to a combination of those with $j_2=1$. This statement can be proved by the following set of transformations of a 5-point conformal block:

\begin{picture}(400,600)(-220,-530)
\put(135,20){\mbox{
$\sum\limits_{t} S_{s,t}\left(\begin{array}{cc} p& j_3 \\ j_1 & j_4\end{array}\right)\ \times$}}
\put(188,-80){\line(0,1){30}}
\put(188,-80){\line(-1,-1){20}}
\put(188,-80){\line(1,-1){20}}
\put(188,-50){\line(-1,1){20}}
\put(188,-50){\line(1,1){30}}
\put(168,-30){\line(-2,1){20}}
\put(168,-30){\line(-1,2){10}}
\put(162,-10){\mbox{$1$}}
\put(223,-15){\mbox{$j_3$}}
\put(166,-47){\mbox{$p$}}
\put(143,-30){\mbox{$r$}}
\put(194,-70){\mbox{$t$}}
\put(158,-110){\mbox{$j_1$}}
\put(208,-110){\mbox{$j_4$}}
\put(-50,0){\line(1,0){60}}
\put(-50,0){\line(-1,1){20}}
\put(-50,0){\line(-1,-1){30}}
\put(10,0){\line(1,1){30}}
\put(10,0){\line(1,-1){30}}
\put(-70,20){\line(-2,1){20}}
\put(-70,20){\line(-1,2){10}}
\put(-95,22){\mbox{$r$}}
\put(-75,38){\mbox{$1$}}
\put(-73,6){\mbox{$p$}}
\put(-95,-30){\mbox{$j_1$}}
\put(-22,7){\mbox{$s$}}
\put(45,30){\mbox{$j_3$}}
\put(45,-30){\mbox{$j_4$}}
\put(-220,-105){\mbox{
$\sum\limits_k  S_{p,k}\left(\begin{array}{cc}1& s \\ r & j_1\end{array}\right)\ \times$}}
\put(-50,-100){\line(1,0){60}}
\put(-50,-100){\line(-1,1){30}}
\put(-50,-100){\line(-1,-1){30}}
\put(10,-100){\line(1,1){30}}
\put(10,-100){\line(1,-1){30}}
\put(-95,-70){\mbox{$r$}}
\put(-95,-130){\mbox{$j_1$}}
\put(-12,-95){\mbox{$s$}}
\put(45,-70){\mbox{$j_3$}}
\put(45,-130){\mbox{$j_4$}}
\put(-30,-100){\line(0,1){30}}
\put(-43,-95){\mbox{$k$}}
\put(-25,-72){\mbox{$1$}}
\put(-220,-225){\mbox{
$\sum\limits_m  S_{s,m}\left(\begin{array}{cc}1& j_3 \\ k & j_4\end{array}\right)\ \times$}}
\put(-50,-220){\line(1,0){60}}
\put(-50,-220){\line(-1,1){30}}
\put(-50,-220){\line(-1,-1){30}}
\put(10,-220){\line(1,-1){30}}
\put(-95,-190){\mbox{$r$}}
\put(-95,-250){\mbox{$j_1$}}
\put(-22,-215){\mbox{$k$}}
\put(45,-250){\mbox{$j_4$}}
\put(10,-220){\line(0,1){20}}
\put(10,-200){\line(-1,1){30}}
\put(10,-200){\line(1,1){30}}
\put(17,-208){\mbox{$m$}}
\put(-25,-180){\mbox{$1$}}
\put(45,-180){\mbox{$j_3$}}
\put(-220,-340){\mbox{
$\sum\limits_t  S_{k,t}\left(\begin{array}{cc}r& m \\ j_1 & j_4\end{array}\right)\ \times$}}
\put(-20,-350){\line(0,1){30}}
\put(-20,-350){\line(-1,-1){20}}
\put(-20,-350){\line(1,-1){20}}
\put(-20,-320){\line(-1,1){20}}
\put(-20,-320){\line(1,2){10}}
\put(-10,-300){\line(-1,2){10}}
\put(-10,-300){\line(1,1){20}}
\put(-29,-280){\mbox{$1$}}
\put(15,-280){\mbox{$j_3$}}
\put(-8,-313){\mbox{$m$}}
\put(-48,-305){\mbox{$r$}}
\put(-14,-340){\mbox{$t$}}
\put(-50,-380){\mbox{$j_1$}}
\put(0,-380){\mbox{$j_4$}}
\put(-220,-470){\mbox{
$\sum\limits_{q} S_{m,p}\left(\begin{array}{cc}r& 1 \\ t & j_3\end{array}\right)\ \times$}}
\put(-20,-480){\line(0,1){30}}
\put(-20,-480){\line(-1,-1){20}}
\put(-20,-480){\line(1,-1){20}}
\put(-20,-450){\line(-1,1){20}}
\put(-20,-450){\line(1,1){30}}
\put(-40,-430){\line(-2,1){20}}
\put(-40,-430){\line(-1,2){10}}
\put(-46,-410){\mbox{$1$}}
\put(15,-415){\mbox{$j_3$}}
\put(-42,-447){\mbox{$q$}}
\put(-65,-430){\mbox{$r$}}
\put(-14,-470){\mbox{$t$}}
\put(-50,-510){\mbox{$j_1$}}
\put(0,-510){\mbox{$j_4$}}
\end{picture}

\noindent Here the initial ``conformal block'' (in the center top of the diagram) can be transformed into the one on the bottom of the diagram either through a four-step chain of $S$-transformations shown in the center, or a one-step $S$-transformation in the right column. At each step the conformal block is replaced by a sum of the $S$-transformed ones, in the dual channel, so the two ways of expanding of the original conformal block correspond respectively to a product of four $S$-matrices shown in the left column, or to a single $S$-matrix in the right column. Note that representations $p$ and $q$ in the diagram can only take values $r\pm 1$. Projecting on the conformal block with either of this values one obtains
\be
S_{s,t}\left(\begin{array}{cc} r\pm1& j_3 \\j_1&j_4\end{array}\right)
= \sum_{k,m} S_{r\pm1,k}\left(\begin{array}{cc} 1& s \\r & j_1\end{array}\right)
S_{s,m}\left(\begin{array}{cc} 1& j_3 \\  k & j_4\end{array}\right)
S_{k,t}\left(\begin{array}{cc} r& m \\j_1&j_4\end{array}\right)
S_{m,r\pm1}\left(\begin{array}{cc} r& 1 \\ t & j_3\end{array}\right)
\label{Sr+1}
\ee
The sums here are over representations $k$ and $m$,
restricted by the following conditions:
\be
k \in 1\otimes s = s\pm 1, \ \ \ &{\rm and} \ \ \
& k\in r\otimes j_1 \ \stackrel{r=1}{=} \ j_1\pm 1, \nn \\
m \in r \otimes t \ \stackrel{r=1}{=} t \pm 1, \ \ \  &{\rm and} \ \ \
& m \in 1\otimes j_3 = j_3\pm 1\,
\label{restr}
\ee
Notice that the relation that we derived here is the same as the pentagon identity~(\ref{pentagon}). This recursion formula is also related to the ``cabling" procedure ($[r]\otimes [1]=[r+1]\oplus [r-1]$) used in \cite{cab} for the derivation of the Racah matrices.

For the four first symmetric representations $[2]$ of $SU_q(2)$ the above procedure gives
\be
S_{st}\left(\begin{array}{cc} 2& 2 \\ 2&2\end{array}\right)
\ \stackrel{(\ref{Sr+1})}{=}\ \sum_{k,m}
S_{2\,k}\left(\begin{array}{cc} 1& s \\ 1&2\end{array}\right)
S_{sm}\left(\begin{array}{cc} 1& 2 \\ k&2\end{array}\right)
S_{kt}\left(\begin{array}{cc} 1& m \\ 2&2\end{array}\right)
S_{m\,2}\left(\begin{array}{cc} 1& 1 \\ t&2\end{array}\right)
\label{S2222}
\ee
In this case each of the indices $s,t$ can take three values $0,2,4$, while $k$ and $l$ are
more restricted by (\ref{restr}):
\be
\begin{array}{c|ccc}
  & t=0 & t=2 & t=4 \\
  & & &  \\
  \hline
  & & & \\
s=0  & \begin{array}{c} k=1 \\ m=1 \end{array}
& \begin{array}{c} k=1 \\ m=1,3 \end{array}
& \begin{array}{c} k=1 \\ m=3 \end{array} \\
 &&&\\
  \hline
  &&&\\
s=2 & \begin{array}{c} k=1,3 \\ m=1 \end{array}
& \begin{array}{c} k=1,3 \\ m=1,3 \end{array}
& \begin{array}{c} k=1,3 \\ m=3 \end{array} \\
 &&&\\
  \hline
  &&&\\
s=4 & \begin{array}{c} k=3 \\ m=1 \end{array}
& \begin{array}{c} k=3 \\ m=1,3 \end{array}
& \begin{array}{c} k=3 \\ m=3 \end{array} \\
 &&&
\end{array}
\ee

The Racah matrices of type~(\ref{RacahwFund}), containing one fundamental representation, can then be calculated with the methods described in the previous section. Indeed, in the $SU_q(2)$ case the degenerate Virasoro conformal block with one fundamental representation $j_2=1$ is described by a hypergeometric function
\be
{}_2F_1 \left(
2+\frac{b^{-2}}{2}(3+j_1+j_3+j_4),1+\frac{b^{-2}}{2}(1+j_1+j_3-j_4); 2+b^{-2}(1+j_1)\ \Big| \ z\,\right)\,
\ee
It leads to the following crossing matrix
\be
M=
\left(
\begin{array}{cc}
M_{-1,-1} & M_{-1,1} \\ M_{1,-1} & M_{11}
\end{array}\right)=
\left(
\begin{array}{cc}
    \frac{\Gamma \left(\frac{j_1+1}{b^2}+2\right) \Gamma \left(-\frac{b^2+j_3+1}{b^2}\right)}{\Gamma \left(\frac{2 b^2+j_1-j_3+j_4+1}{2 b^2}\right) \Gamma \left(-\frac{-j_1+j_3+j_4+1}{2 b^2}\right)} & \frac{\Gamma \left(\frac{j_1+1}{b^2}+2\right) \Gamma \left(\frac{b^2+j_3+1}{b^2}\right)}{\Gamma \left(\frac{2 b^2+j_1+j_3-j_4+1}{2 b^2}\right) \Gamma \left(\frac{4 b^2+j_1+j_3+j_4+3}{2 b^2}\right)} \\
    \frac{\Gamma \left(-\frac{j_1+1}{b^2}\right) \Gamma \left(-\frac{b^2+j_3+1}{b^2}\right)}{\Gamma \left(-\frac{j_1+j_3-j_4+1}{2 b^2}\right) \Gamma \left(-\frac{2 b^2+j_1+j_3+j_4+3}{2 b^2}\right)} & \frac{\Gamma \left(-\frac{j_1+1}{b^2}\right) \Gamma \left(\frac{b^2+j_3+1}{b^2}\right)}{\Gamma \left(-\frac{j_1-j_3+j_4+1}{2 b^2}\right) \Gamma \left(\frac{2 b^2-j_1+j_3+j_4+1}{2 b^2}\right)} \\
\end{array}
\right)
\ee
where matrix $M$ is related to the $S$-matrix through the relation
\be
 S_{j_s\; j_t}\left[\begin{array}{cc}
    1 & j_3\\
    j_1 & j_4\\
 \end{array}\right]=\sum\limits_{h,h'=\pm 1}\delta(j_s-j_1-h)\delta(j_t-j_3-h')
 M_{h,h'}(j_1,j_3,j_4)\,
\label{SviaM}
\ee

\section{Elementary examples of link diagrams in plat representation}
\label{sec:examples}

In the previous section we have explained how the operators $S$ and $T$ satisfying (\ref{STrels0}) naturally arise
as modular transformations of the conformal blocks. Using this representation of $S$ and $T$ one can construct link diagrams in the braid, plat or other presentations. In this section we will consider the application of the approach to simple families of knots and links.

It is somewhat convenient to use the TQFT language in construction of knot invariants. For example, in the plat presentation the link diagram can be cut in three pieces: the initial state given by $n$ cups, a braid on $2n$ strands representing the evolution of the initial state and the contraction with (projection onto) the final state given by $n$ inverted cups. This is described by the following formula
\be
\langle\underbrace{\cup\ldots\cup}_{n}|\,{\mathcal{B}}^{(2n)}\,|\underbrace{\cup\ldots\cup}_n\rangle\,
\ee
If the given diagram represents the minimum possible $n$ such a knot or link is called $n$-bridge.

\subsection{One-bridge case}

The link diagram in this case is the initial state represented by a cup (bridge), the final state given by inverted cup and two strands connecting the two cups with an arbitrary number of intertwinings between the strands. The result is trivial: independently of the number of intertwinings the diagram always represents the unknot. For calculating the knot invariant one needs to fix the relation between the TQFT state $|\cup\rangle$ and a vector in the Hilbert space associated with conformal blocks. This can be done with the use of a 3-point function (conformal block) with one external leg carrying the trivial representation, while the other two carrying the representation of the unknot $R$. In other words, we use a map
\be
|\cup\rangle \longrightarrow {\rm Y_0}\,
\ee
where ${\rm Y_0}$ denotes the conformal block. It is natural to normalize the TQFT state so that it yields the unknot invariant:
\be
\langle\cup|\cup\rangle = \dim_q R\,
\ee
while the conformal blocks are defined to be orthonormal $\langle{\rm Y_0}|{\rm Y_0}\rangle=1$. Thus formally we set
\be
|\cup\rangle = \sqrt{\dim_q R}\,|{\rm Y_0}\rangle\,
\ee
Then for $n$ intertwinings, the invariant is given by
\be
\langle{\rm Y_0}|T^{n}|{\rm Y_0}\rangle=\lambda_0^{n}\dim_q R=q^{n}\dim_q R\,
\ee
where $\lambda_0=q$ is the eigenvalue of the $T$-matrix in the trivial representation.

\subsection{Two-bridge case}

In this case the initial state is $|\cup\cup\,\rangle$ and the associated conformal block $|{\rm Y\!\__0\!\_Y}\rangle$ is shown on the figure below (lower line):

\begin{picture}(100,190)(-50,-80)
\put(0,0){\vector(0,-1){20}}
\put(20,-20){\vector(0,1){20}}
\qbezier(0,-20)(0,-30)(10,-30)
\qbezier(20,-20)(20,-30)(10,-30)
\put(80,0){\vector(0,-1){20}}
\put(60,-20){\vector(0,1){20}}
\qbezier(60,-20)(60,-30)(70,-30)
\qbezier(80,-20)(80,-30)(70,-30)
\put(10,-50){\line(1,0){60}}
\put(10,-50){\line(-2,1){20}}
\put(10,-50){\line(-2,-1){20}}
\put(70,-50){\line(2,1){20}}
\put(70,-50){\line(2,-1){20}}
\put(38,-47){\mbox{$0$}}
\qbezier(0,0)(0,15)(10,20)
\qbezier(10,20)(20,25)(20,40)
\qbezier(20,0)(20,15)(10,20)
\qbezier(10,20)(0,25)(0,40)
\put(10,60){\line(1,0){60}}
\put(10,60){\line(-2,1){20}}
\put(10,60){\line(-2,-1){20}}
\put(70,60){\line(2,1){20}}
\put(70,60){\line(2,-1){20}}
\put(38,63){\mbox{$0$}}
\qbezier(-18,68)(-22,60)(-18,52)
\put(-18.5,67){\vector(1,2){2}}
\put(-18.5,53){\vector(1,-2){2}}
\put(200,0){\vector(0,-1){20}}
\put(220,-20){\vector(0,1){20}}
\qbezier(200,-20)(200,-30)(210,-30)
\qbezier(220,-20)(220,-30)(210,-30)
\put(300,0){\vector(0,-1){20}}
\put(280,-20){\vector(0,1){20}}
\qbezier(280,-20)(280,-30)(290,-30)
\qbezier(300,-20)(300,-30)(290,-30)
\put(210,-50){\line(1,0){80}}
\put(210,-50){\line(-2,1){20}}
\put(210,-50){\line(-2,-1){20}}
\put(290,-50){\line(2,1){20}}
\put(290,-50){\line(2,-1){20}}
\put(248,-47){\mbox{$0$}}
\qbezier(220,0)(220,15)(230,20)
\qbezier(230,20)(240,25)(240,40)
\qbezier(280,0)(280,15)(270,20)
\qbezier(270,20)(260,25)(260,40)
\put(250,60){\line(0,1){20}}
\put(250,60){\line(-2,-1){20}}
\put(250,60){\line(2,-1){20}}
\put(250,80){\line(-2,1){20}}
\put(250,80){\line(2,1){20}}
\put(254,67){\mbox{$j$}}
%
%
\put(-12,-25){\mbox{$\bar r_1$}}
\put(23,-25){\mbox{$r_1$}}
\put(48,-25){\mbox{$r_2$}}
\put(83,-25){\mbox{$\bar r_2$}}
\put(-22,-65){\mbox{$\bar r_1$}}
\put(-22,-42){\mbox{$r_1$}}
\put(95,-42){\mbox{$r_2$}}
\put(95,-65){\mbox{$\bar r_2$}}
\put(188,-25){\mbox{$\bar r_1$}}
\put(223,-25){\mbox{$r_1$}}
\put(268,-25){\mbox{$ r_2$}}
\put(303,-25){\mbox{$\bar r_2$}}
\put(178,-65){\mbox{$\bar r_1$}}
\put(178,-42){\mbox{$r_1$}}
\put(315,-42){\mbox{$r_2$}}
\put(315,-65){\mbox{$\bar r_2$}}
\put(215,50){\mbox{$\bar r_1$}}
\put(215,90){\mbox{$r_1$}}
\put(275,90){\mbox{$r_2$}}
\put(275,50){\mbox{$\bar r_2$}}
\put(200,70){\mbox{$\bigoplus_j$}}
\end{picture}

One can apply the $T$ matrix~(\ref{Tgen}) directly to braid the strands in the 12 and 34 channels (numbers $1,2,3,4$ here
label the vertical lines in the above figure from left to right). For the example shown in the figure one has
\be
T\otimes I^{\otimes 2}\,|{\rm Y\!\__0\!\_Y}\rangle = \lambda_0(\bar r_1,r_1)|{\rm Y\!\__0\!\_Y}\rangle\,
\ee
where again $\lambda_0$ is the eigenvalue of the matrix $T$ corresponding to the trivial representation. In order to braid the strands in the channel 23 one has to bring the points $2$ and $3$ close to each other, that is to make the $S$ transformation changing the conformal block basis (right part of the figure). One way to do this is to first expand the original conformal block in the dual basis:
\be
\label{changeofbasis}
|{\rm Y\!\__0\!\_Y}\rangle = \sum\limits_j S_{0j}\left[\begin{array}{cc} r_1 & r_2 \\ \bar{r}_1 & \bar r_2\end{array}\right]|{\rm I\!\_Y_j\!\_I}\rangle\,
\ee
Then $T$ acts diagonally on the states of the $|{\rm I\!\_Y_j\!\_I}\rangle$ basis:
\be
I\otimes T \otimes I|{\rm I\!\_Y_j\!\_I}\rangle = \lambda_j(r_1,r_2)|{\rm I\!\_Y_j\!\_I}\rangle\,
\ee
The invariant is obtained by contracting the result of the evolution with the conjugate of the initial state using the normalization
\be
\langle{\rm Y\!\__i\!\_Y}|{\rm Y\!\__j\!\_Y}\rangle = \delta_{ij}\,
\ee

The family of the knots/links that is obtained as the result of this procedure includes 2-strand links and knots, twist knots, antiparallel 2-strand links, double braids from \cite{evo} and, more generally, all two-bridge knots/links.

\paragraph{Two unknots.} If no $S$ operators are applied, we get two disconnected unknots.
The answer for two fundamental representations
\be
\lambda_0^{n_1}\lambda_{0}^{n_2} = q^{n_1+n_2}\,
\ee
Up to the framing factor this is the fully reduced knot polynomial
(i.e. unreduced expression $[2]^2$ is divided by a square of the quantum dimension $[2]$).

\paragraph{2-strand torus links.} The plat diagram and the sequence of modular transformations in this case are:

\begin{picture}(300,280)(-50,-50)
\put(0,0){\vector(0,-1){20}}
\put(20,-20){\vector(0,1){20}}
\qbezier(0,-20)(0,-30)(10,-30)
\qbezier(20,-20)(20,-30)(10,-30)
\put(100,0){\vector(0,-1){20}}
\put(80,-20){\vector(0,1){20}}
\qbezier(80,-20)(80,-30)(90,-30)
\qbezier(100,-20)(100,-30)(90,-30)
\qbezier(20,0)(20,15)(30,20)
\qbezier(30,20)(40,25)(40,40)
\qbezier(80,0)(80,15)(70,20)
\qbezier(70,20)(60,25)(60,40)
\qbezier(40,40)(40,55)(50,60)
\qbezier(50,60)(60,65)(60,80)
\qbezier(60,40)(60,55)(50,60)
\qbezier(50,60)(40,65)(40,80)
\put(43,90){\mbox{$\ldots$}}
\qbezier(40,100)(40,115)(50,120)
\qbezier(50,120)(60,125)(60,140)
\qbezier(60,100)(60,115)(50,120)
\qbezier(50,120)(40,125)(40,140)
\qbezier(20,180)(20,165)(30,160)
\qbezier(30,160)(40,155)(40,140)
\qbezier(80,180)(80,165)(70,160)
\qbezier(70,160)(60,155)(60,140)
\put(0,200){\vector(0,-1){220}}
\put(20,180){\vector(0,1){20}}
\qbezier(0,200)(0,210)(10,210)
\qbezier(20,200)(20,210)(10,210)
\put(100,200){\vector(0,-1){220}}
\put(80,180){\vector(0,1){20}}
\qbezier(80,200)(80,210)(90,210)
\qbezier(100,200)(100,210)(90,210)
\put(210,-15){\line(1,0){80}}
\put(210,-15){\line(-2,1){20}}
\put(210,-15){\line(-2,-1){20}}
\put(290,-15){\line(2,1){20}}
\put(290,-15){\line(2,-1){20}}
\put(248,-10){\mbox{$0$}}
\put(210,195){\line(1,0){80}}
\put(210,195){\line(-2,1){20}}
\put(210,195){\line(-2,-1){20}}
\put(290,195){\line(2,1){20}}
\put(290,195){\line(2,-1){20}}
\put(248,200){\mbox{$0$}}

\put(250,60){\line(0,1){20}}
\put(250,60){\line(-2,-1){20}}
\put(250,60){\line(2,-1){20}}
\put(250,80){\line(-2,1){20}}
\put(250,80){\line(2,1){20}}
\put(254,67){\mbox{$j$}}
\qbezier(234,95)(250,110)(266,95)
\put(236,97){\vector(-1,-1){2}}
\put(264,97){\vector(1,-1){2}}
\put(246,110){\mbox{$T_j^{2k}$}}
\put(-12,-25){\mbox{$\bar r_1$}}
\put(23,-25){\mbox{$r_1$}}
\put(68,-25){\mbox{$r_2$}}
\put(103,-25){\mbox{$\bar r_2$}}
%
%
%
\put(178,-30){\mbox{$\bar r_1$}}
\put(178,-7){\mbox{$r_1$}}
\put(315,-7){\mbox{$r_2$}}
\put(315,-30){\mbox{$\bar r_2$}}
\put(215,50){\mbox{$\bar r_1$}}
\put(215,90){\mbox{$r_1$}}
\put(275,90){\mbox{$r_2$}}
\put(275,50){\mbox{$\bar r_2$}}
\put(200,70){\mbox{$\bigoplus_j$}}
\put(178,208){\mbox{$r_1$}}
\put(178,180){\mbox{$\bar r_1$}}
\put(315,180){\mbox{$\bar r_2$}}
\put(315,208){\mbox{$  r_2$}}
\put(250,10){\vector(0,1){30}}
\put(250,140){\vector(0,1){30}}
\put(257,20){\mbox{$S_{0j}$}}
\put(257,150){\mbox{$S_{j0}$}}
\end{picture}

The corresponding analytical expression is
\be
\label{2strtlinkgen}
\langle{\rm Y\!\__0\!\_Y}|(STS^\dagger)^{2k}|{\rm Y\!\__0\!\_Y}\rangle=\sum_j S_{j0}\left[\begin{array}{cc}r_2 & \bar r_2\\ r_1& \bar r_1\end{array}\right]
\ \lambda_j\big[r_1,r_2\big]^{2k}\
S_{0j}\left[\begin{array}{cc} r_1 & r_2 \\  \bar r_1 & \bar r_2 \end{array}\right]\,
\ee
where $\beta(b_2^{2k})=(STS^\dagger)^{2k}=ST^{2k}S^\dagger$ is the representation of the corresponding braid. Notice that due to an obvious symmetry the following property under the permutations of representations must hold
\be
S^{\dagger}_{ij}\left[\begin{array}{cc} \bar r_2 & \bar r_1 \\  r_2 &  r_1 \end{array}\right] = S^{-1}_{ij}\left[\begin{array}{cc} \bar r_2 & \bar r_1 \\  r_2 &
 r_1 \end{array}\right]=\pm S_{ij}\left[\begin{array}{cc}r_2 & \bar r_2\\ r_1& \bar r_1\end{array}\right]\,
\ee

To illustrate the method and our conventions let us go carefully through the steps outlined by the above diagram and derive formula~(\ref{2strtlinkgen}). To braid in the $23$-channel we first expand over the dual basis, where the generator $b_2$ is diagonal:
\be
b_2^{2k}:\quad |{\rm Y\!\__0\!\_Y}\rangle = \sum\limits_{j}S_{0j}\left[\begin{array}{cc} r_1 & r_2 \\ \bar r_1 & \bar r_2 \end{array}\right]|{\rm I\!\_Y_j\!\_I}\rangle \longrightarrow \sum\limits_{j}\lambda_j\big[r_1,r_2\big]^{2k}S_{0j}\left[\begin{array}{cc} r_1 &  r_2 \\  \bar r_1 & \bar r_2 \end{array}\right]|{\rm I\!\_Y_j\!\_I}\rangle
\ee
To contract the result of the evolution with the conjugate of the initial state, we need another $S$-transformation. Equivalently, one may use the orthonormality of conformal blocks and an equation like~(\ref{changeofbasis}) to derive the scalar product directly
\be
\langle{\rm Y\!\__0\!\_Y}|{\rm I\!\_Y_j\!\_I}\rangle = S_{j0}\left[\begin{array}{cc}r_2 & \bar r_2\\ r_1& \bar r_1\end{array}\right]
\ee
which completes the derivation.

In fact, formula~(\ref{2strtlinkgen}) can be obtained as the element of the matrix $ST^{2k}S^\dagger$,
\be
\label{2strlinkgen2}
(ST^{2k}S^\dagger)_{ij}=\langle{\rm Y\!\__i\!\_Y}|ST^{2k}S^\dagger|{\rm Y\!\__j\!\_Y}\rangle
\ee
with $i$ and $j$ corresponding to the trivial representation. In the case of two fundamental representations, $r_1=r_2=[1]$, of $SU_q(2)$ we can use matrices (\ref{ST2}) and obtain:
\be
ST^{2k}S^\dagger\ \stackrel{(\ref{ST2})}{=}
\ \frac{1}{[2]^2}\left(\begin{array}{cc} q^{2k}+q^{-2k}[3] & \big(q^{2k}-q^{-2k}\big)\sqrt{[3]} \\ \\
\big(q^{2k}-q^{-2k}\big)\sqrt{[3]} & q^{2k}[3]+q^{-2k} \end{array}\right)
\label{2straN2}
\ee
Expression in the upper left
corner (the matrix element $_{11}$)
is exactly the {\it unreduced} Jones polynomial
\be
J_{[1],[1]}^{[2,2k]} = \left.\left(q^{2k}\frac{[N][N-1]}{[2]}+q^{-2k}\frac{[N][N+1]}{[2]}\right)
\right|_{N=2} = [3]q^{-2k}+q^{2k}
\label{2strlinkJones}
\ee
for the 2-strand torus links (in the Rosso-Jones, rather than topological framing).
We also notice that the matrix element $_{22}$ gives the same result up to the change $q\to -1/q$, that is it yields the polynomial of the mirror image of the knot.

\paragraph{2-strand torus knots.}

The only difference in this case is that even power $2k$ is substituted by odd $2k+1$, which is only possible for the coincident representations $r_1=r_2$. Indeed, after an odd number of intertwinings in the middle channel one has to replace the top diagram on the right part of the above picture with
\begin{picture}(100,50)(-150,-75)
\put(10,-50){\line(1,0){60}}
\put(10,-50){\line(-2,1){20}}
\put(10,-50){\line(-2,-1){20}}
\put(70,-50){\line(2,1){20}}
\put(70,-50){\line(2,-1){20}}
\put(38,-47){\mbox{$0$}}
\put(-22,-65){\mbox{$\bar r_1$}}
\put(-22,-42){\mbox{$r_2$}}
\put(95,-42){\mbox{$r_1$}}
\put(95,-65){\mbox{$\bar r_2$}}
\end{picture}

\noindent
This is only possible (the singlet in the intermediate line) if $r_1=r_2$.

As to formula (\ref{2straN2}), it remains just the same, with the obvious change
$2k\longrightarrow 2k+1$, and the upper left $_{11}$ element of the matrix
reproduces the reduced Jones polynomial
\be
ST^{2k+1}S \ \stackrel{(\ref{ST2})}{=}
\ \frac{1}{[2]^2}\left(\begin{array}{cc} q^{2k+1}-q^{-2k-1}[3] & (q^{2k+1}-q^{-2k-1})\sqrt{[3]} \\ \\
(q^{2k+1}-q^{-2k-1})\sqrt{[3]} & q^{2k+1}[3]-q^{-2k-1} \end{array}\right) =
\ \left(\begin{array}{ccc} \frac{1}{[2]^2}J_{[1]}^{[2,2k+1]} && \ldots \\ \\ \ldots && \ldots
 \end{array}\right)
\label{2straN2kn}
\ee
where
\be
J_{[1]}^{[2,2k+1]} = \left.\left(q^{2k+1}\,\frac{[N][N-1]}{[2]}-q^{-2k-1}\,\frac{[N][N+1]}{[2]}\right)
\right|_{N=2} = \Big(q^{2k+1}-q^{-2k-1}[3]
\ee
Note that, like in (\ref{2strlinkJones}),
the Jones polynomial appeared in the Rosso-Jones rather than topological framing. Again, the matrix element $_{22}$ gives the polynomial of the mirror knot, $q\to -1/q$.

\paragraph{Twist knots} differ from the above examples in two ways. First, there are two additional twists in the $12$-channel. Second, the braid is closed in a different manner:

\begin{picture}(300,380)(-50,-40)
\put(0,0){\vector(0,-1){20}}
\put(20,-20){\vector(0,1){20}}
\qbezier(0,-20)(0,-30)(10,-30)
\qbezier(20,-20)(20,-30)(10,-30)
\put(80,0){\vector(0,-1){20}}
\put(100,-20){\vector(0,1){20}}
\qbezier(80,-20)(80,-30)(90,-30)
\qbezier(100,-20)(100,-30)(90,-30)
\qbezier(20,0)(20,15)(30,20)
\qbezier(30,20)(40,25)(40,40)
\qbezier(80,0)(80,15)(70,20)
\qbezier(70,20)(60,25)(60,40)
\qbezier(40,40)(40,55)(50,60)
\qbezier(50,60)(60,65)(60,80)
\qbezier(60,40)(60,55)(50,60)
\qbezier(50,60)(40,65)(40,80)
\put(43,90){\mbox{$\ldots$}}
\qbezier(40,100)(40,115)(50,120)
\qbezier(50,120)(60,125)(60,140)
\qbezier(60,100)(60,115)(50,120)
\qbezier(50,120)(40,125)(40,140)
\qbezier(20,180)(20,165)(30,160)
\qbezier(30,160)(40,155)(40,140)
%
%
\put(0,310){\vector(0,-1){53}}
\put(20,260){\vector(0,1){2}}
\qbezier(40,300)(40,310)(50,310)
\qbezier(60,300)(60,310)(50,310)
\put(60,300){\vector(0,-1){160}}
%
\qbezier(0,310)(0,320)(10,320)
\qbezier(100,310)(100,320)(90,320)
\put(90,320){\vector(-1,0){80}}
\put(100,310){\line(0,-1){310}}
\qbezier(0,180)(0,195)(10,200)
\qbezier(10,200)(20,205)(20,220)
\qbezier(20,180)(20,195)(10,200)
\qbezier(10,200)(0,205)(0,220)
\qbezier(0,220)(0,235)(10,240)
\qbezier(10,240)(20,245)(20,260)
\qbezier(20,220)(20,235)(10,240)
\qbezier(10,240)(0,245)(0,260)
\put(0,180){\vector(0,-1){200}}
\qbezier(40,300)(40,285)(30,280)
\qbezier(30,280)(20,275)(20,260)

\put(210,-15){\line(1,0){80}}
\put(210,-15){\line(-2,1){20}}
\put(210,-15){\line(-2,-1){20}}
\put(290,-15){\line(2,1){20}}
\put(290,-15){\line(2,-1){20}}
\put(248,-10){\mbox{$0$}}
\put(210,215){\line(1,0){80}}
\put(210,215){\line(-2,1){20}}
\put(210,215){\line(-2,-1){20}}
\put(290,215){\line(2,1){20}}
\put(290,215){\line(2,-1){20}}
\put(248,220){\mbox{$l$}}
\qbezier(186,210)(182,215)(186,220)
\put(186,220){\vector(1,2){2}}
\put(186,210){\vector(1,-2){2}}
\put(165,212){\mbox{$T_l^2$}}
\put(135,212){\mbox{$\bigoplus_{l}$}}
\put(250,60){\line(0,1){20}}
\put(250,60){\line(-2,-1){20}}
\put(250,60){\line(2,-1){20}}
\put(250,80){\line(-2,1){20}}
\put(250,80){\line(2,1){20}}
\put(254,67){\mbox{$j$}}
\qbezier(234,95)(250,110)(266,95)
\put(236,97){\vector(-1,-1){2}}
\put(264,97){\vector(1,-1){2}}
\put(246,110){\mbox{$T_j^{2k}$}}
\put(250,290){\line(0,1){20}}
\put(250,290){\line(-2,-1){20}}
\put(250,290){\line(2,-1){20}}
\put(250,310){\line(-2,1){20}}
\put(250,310){\line(2,1){20}}
\put(254,297){\mbox{$0$}}
%
%
\put(-12,-25){\mbox{$\bar r$}}
\put(23,-25){\mbox{$r$}}
\put(68,-25){\mbox{$\bar r$}}
\put(103,-25){\mbox{$r$}}
%
%
%
\put(182,-30){\mbox{$\bar r$}}
\put(182,-7){\mbox{$r$}}
\put(315,-7){\mbox{$\bar r$}}
\put(315,-30){\mbox{$r$}}
\put(221,50){\mbox{$\bar r$}}
\put(221,90){\mbox{$r$}}
\put(275,90){\mbox{$\bar r$}}
\put(275,50){\mbox{$r$}}
\put(200,70){\mbox{$\bigoplus_j$}}
\put(182,228){\mbox{$r$}}
\put(182,197){\mbox{$\bar r$}}
\put(315,200){\mbox{$ r$}}
\put(315,228){\mbox{$\bar  r$}}
\put(221,280){\mbox{$\bar r$}}
\put(221,320){\mbox{$r$}}
\put(275,320){\mbox{$\bar r$}}
\put(275,280){\mbox{$r$}}
\put(250,10){\vector(0,1){30}}
\put(250,145){\vector(0,1){40}}
\put(250,245){\vector(0,1){20}}
\put(257,20){\mbox{$S_{0j}$}}
\put(257,160){\mbox{$S_{jl}$}}
\put(257,250){\mbox{$S_{l0}$}}
\end{picture}

\noindent so that the final state of the TQFT evolution is a different conformal block. Note that in order to have a closed oriented line we should not change the order in which representation and its conjugate appear in the last two vertical lines.

The analytical expression is now
\be
\langle{\rm I\!\_Y_0\!\_I}|T^2(STS^\dagger)^{2k}|{\rm Y\!\__0\!\_Y}\rangle=\sum_{l,j} S_{l0}\left[\begin{array}{cc}  r & \bar r \\ \bar r & r \end{array}\right] \
\lambda_l\big[r,\bar r\big]^2 \
S_{jl}\left[\begin{array}{cc} \bar r &  r \\  r & \bar r\end{array}\right] \
\lambda_j\big[r,\bar r\big]^{2k} \
S_{0j}\left[\begin{array}{cc}  r &  \bar r \\  \bar r &  r\end{array}\right]
\ee
In the case of the fundamental representation $r=[1]=\overline{[1]}$ of $SU_q(2)$ we can use (\ref{ST2}) and obtain:
\be
ST^2ST^{2k}S \ \stackrel{(\ref{ST2})}{=}
\frac{1}{[2]^2} \left(\begin{array}{ccc} q^{2k-1}(q^2+q^{-2})+q^{-2k}\{q^3\} &&
\Big(q^{2k-1}(q^2+q^{-2})-q^{-2k}\{q\}\Big)\sqrt{[3]} \\ \\
\Big(q^{2k}\{q\}+q^{1-2k}(q^2+q^{-2})\Big)\sqrt{[3]} && q^{2k}\{q^3\}-q^{1-2k}(q^2+q^{-2}) \end{array}\right)
= \nn \\ =
\ \left(\begin{array}{ccc} \ldots && \ldots \\ \\ \ldots &&
-\frac{q^{-2k-2}}{[2]}J_{[1]}^{Tw(k)} \end{array}\right)
\label{STtw2}
\ee
Again, the diagonal elements are related by $q\to -1/q$. In the conventions used by~\cite{GMMlast} it is the lower right element of the matrix that is related to the (unreduced) Jones polynomials of the twist knots in the topological framing:
\be
J_{[1]}^{Tw(k)} = [2]\left(\left. 1+ \frac{A^{k+1}\{A^{-k}\}}{\{A\}}\{Aq\}\{A/q\}\right|_{A=q^2}\right)
= -q^{4k+2}\{q^3\}+ q^3(q^2+q^{-2})
= \nn \\
= -q^{2k+2}\Big( q^{2k}\{q^3\} - q^{1-2k}(q^2+q^{-2})\Big)
\ee

\paragraph{Other families.}
Matrix multiplication allows one to write a direct generalization of~(\ref{STtw2}) to other knot series:
\be
\label{OtherFam1}
(ST^{2k_2}ST^{2k_1}S)_{11} \ \stackrel{(\ref{ST2})}{=} \ \frac{1}{[2]^3}\left([3] \left(q^{4 k_1}+q^{4 k_2}-1\right) q^{-2 (k_1+k_2)}+q^{2 (k_1+k_2)}\right),
\ee
\be
\label{OtherFam2}
(ST^{2k_2+1}ST^{2k_1+1}S)_{11} \ \stackrel{(\ref{ST2})}{=}\ \frac{1}{[2]^3}\left([3] \left(-q^{4 k_1+2}-q^{4 k_2+2}-1\right) q^{-2 \left(k_1+k_2+1\right)}+q^{2 \left(k_1+k_2+1\right)}\right),
\ee
\be
\label{OtherFam3}
(ST^{2k_3}ST^{2k_2}ST^{2k_1}S)_{11} \ \stackrel{(\ref{ST2})}{=} \ \frac{1}{[2]^4}\left([3] \left(-q^{4 k_1}+q^{4 \left(k_1+k_2\right)}-q^{4 k_3}+q^{4 \left(k_1+k_3\right)}+q^{4 \left(k_2+k_3\right)}+1\right) q^{-2 \left(k_1+k_2+k_3\right)} + \right. \nn
\\ \left. +[3]^2 q^{4 k_2-2 \left(k_1+k_2+k_3\right)}+q^{2 \left(k_1+k_2+k_3\right)}\right),
\ee
and so on.

This result can be compared with the formulae for the corresponding series. Indeed, equations~(\ref{OtherFam1}) and~(\ref{OtherFam2}) are symmetric with respect to $k_1\leftrightarrow k_2$.  For equation~(\ref{OtherFam1})
\begin{itemize}
\item for either $k_1,k_2=1$ we recover back the twist series;

\item for $k_1=-k_2=2$, or $k_1=-k_2=-2$ one gets the $8_3$ knot from the Rolfsen table and the relation to the corresponding Jones polynomial (\texttt{katlas.org})
\be
(\ref{OtherFam1})\Big|_{k_1=-k_2=2} \ {=}\ \frac{1}{[2]}\,J_{[1]}^{\,8_3}
\ee

\item For $k_1=k_2=2$ one gets
\be
(\ref{OtherFam1})\Big|_{k_1=k_2=2}  \ {=}\ \left(\frac{q^8}{[2]}\,J_{[1]}^{\,7_4}\right)_{q\to\frac1q}
\ee

\item For $k_1=2, k_2=3$, or $k_1=3, k_2=2$ one gets
\be
(\ref{OtherFam1})\Big|_{k_1=2,k_2=3}  \ {=}\ \left(\frac{q^{10}}{[2]}\,J_{[1]}^{\,9_5}\right)_{q\to\frac1q}
\ee

\end{itemize}

For equation~(\ref{OtherFam2})
\begin{itemize}

\item for $k_1=k_2=0$ we get a product of two unknots,
\be
(\ref{OtherFam2})\Big|_{k_1=k_2=0}  \ {=}\ -\,\frac{1}{q}\,\frac{(\dim_q[1])^2}{[2]^2}=-\,\frac{1}{q}\,;
\ee

\item for $k_1=0$, $k_2=-1$, or $k_1=-1$, $k_2=0$ we get the Hopf link,
\be
(\ref{OtherFam2})\Big|_{k_1=0,k_2=-1}  \ {=}\ \left(\frac{1}{q[2]}\,J_{[1]}^{\,2_1^2}\right)_{q\to\frac1q}\,;
\ee

\item for $k_1=0$, $k_2=1$, or $k_1=1$, $k_2=0$ we get the Hopf link in a different framing,
\be
(\ref{OtherFam2})\Big|_{k_1=0,k_2=1}  \ {=}\ \left(\frac{1}{q^3[2]}\,J_{[1]}^{\,2_1^2}\right)_{q\to\frac1q}\,;
\ee

\item for $k_1=k_2=1$
\be
(\ref{OtherFam2})\Big|_{k_1=k_2=1}  \ {=}\ \left(\frac{q^{-3}}{[2]} J_{[1]}^{5_1^2}\right)_{q\to\frac1q}\,;
\ee

\item for $k_1=1$, $k_2=-2$, or $k_1=-1$, $k_2=2$
\be
(\ref{OtherFam2})\Big|_{k_1=1,k_2=-2}  \ {=}\ \left(\frac{q^9}{[2]}\,J_{[1]}^{\,6_2^2}\right)_{q\to\frac1q}\,;
\ee

\end{itemize}

\ldots

\subsection{Three-bridge case}

In the plat presentation, the natural initial state and the corresponding conformal block would be

\begin{picture}(100,120)(-120,-100)
\put(0,0){\vector(0,-1){20}}
\put(20,-20){\vector(0,1){20}}
\qbezier(0,-20)(0,-30)(10,-30)
\qbezier(20,-20)(20,-30)(10,-30)
\put(100,0){\vector(0,-1){20}}
\put(80,-20){\vector(0,1){20}}
\qbezier(80,-20)(80,-30)(90,-30)
\qbezier(100,-20)(100,-30)(90,-30)
\put(160,0){\vector(0,-1){20}}
\put(180,-20){\vector(0,1){20}}
\qbezier(160,-20)(160,-30)(170,-30)
\qbezier(180,-20)(180,-30)(170,-30)
\put(10,-50){\line(1,0){160}}
\put(10,-50){\line(-2,1){20}}
\put(10,-50){\line(-2,-1){20}}
\put(170,-50){\line(2,1){20}}
\put(170,-50){\line(2,-1){20}}
\put(90,-50){\line(0,-1){30}}
\put(90,-80){\line(-2,-1){20}}
\put(90,-80){\line(2,-1){20}}
\put(48,-47){\mbox{$0$}}
\put(128,-47){\mbox{$0$}}
\put(94,-70){\mbox{$0$}}
\put(8,-5){\mbox{$r_l$}}
\put(168,-5){\mbox{$r_r$}}
\put(90,-5){\mbox{$r$}}
\end{picture}

\noindent where the above picture shows one of the possible mutual orientations of the three lines.
However for other examples, which we will consider later, it is more convenient to use another presentation and consequently, another conformal block

\begin{picture}(100,120)(-120,-100)
\put(40,0){\vector(0,-1){10}}
\put(60,-10){\vector(0,1){10}}
\qbezier(40,-10)(40,-20)(50,-20)
\qbezier(60,-10)(60,-20)(50,-20)
\put(140,0){\vector(0,-1){10}}
\put(120,-10){\vector(0,1){10}}
\qbezier(120,-10)(120,-20)(130,-20)
\qbezier(140,-10)(140,-20)(130,-20)
%
%
\put(0,0){\vector(0,-1){20}}
\put(180,-20){\vector(0,1){20}}
\qbezier(0,-20)(0,-30)(10,-30)
\qbezier(180,-20)(180,-30)(170,-30)
\put(10,-30){\vector(1,0){80}}
\put(90,-30){\line(1,0){80}}
\put(10,-50){\line(1,0){160}}
\put(10,-50){\line(-2,-1){20}}
\put(170,-50){\line(2,-1){20}}
\put(50,-50){\line(0,-1){30}}
\put(130,-50){\line(0,-1){30}}
\put(50,-80){\line(-2,-1){20}}
\put(50,-80){\line(2,-1){20}}
\put(130,-80){\line(-2,-1){20}}
\put(130,-80){\line(2,-1){20}}
\put(58,-70){\mbox{$0$}}
\put(118,-70){\mbox{$0$}}
\put(90,-48){\mbox{$r$}}
\end{picture}

Depending on the knot/link it is convenient to use either of the above bases of conformal blocks. Different bases can be connected through the following chain of modular transformations:

\begin{picture}(300,200)(-60,-90)
\put(0,50){\line(1,0){80}}
\put(0,50){\line(-2,1){20}}
\put(0,50){\line(-2,-1){20}}
\put(80,50){\line(2,1){20}}
\put(80,50){\line(2,-1){20}}
\put(40,50){\line(0,1){30}}
\put(40,80){\line(-1,2){10}}
\put(40,80){\line(1,2){10}}
\put(18,40){\mbox{$0$}}
\put(58,40){\mbox{$0$}}
\put(44,62){\mbox{$0$}}
\put(210,50){\line(1,0){120}}
\put(210,50){\line(-2,1){20}}
\put(210,50){\line(-2,-1){20}}
\put(330,50){\line(2,1){20}}
\put(330,50){\line(2,-1){20}}
\put(250,50){\line(-1,2){10}}
\put(290,50){\line(1,2){10}}
\put(228,40){\mbox{$0$}}
\put(308,40){\mbox{$0$}}
\put(268,55){\mbox{$j$}}
\put(250,-50){\line(1,0){40}}
\put(250,-50){\line(-2,-1){20}}
\put(250,-50){\line(-2,1){40}}
\put(290,-50){\line(2,-1){20}}
\put(290,-50){\line(2,1){40}}
\put(230,-40){\line(-1,2){10}}
\put(310,-40){\line(1,2){10}}
\put(267,-46){\mbox{$j$}}
\put(237,-40){\mbox{$k$}}
\put(298,-40){\mbox{$l$}}
\put(40,-50){\line(0,-1){20}}
\put(40,-70){\line(-2,-1){20}}
\put(40,-50){\line(-2,1){40}}
\put(40,-70){\line(2,-1){20}}
\put(40,-50){\line(2,1){40}}
\put(20,-40){\line(-1,2){10}}
\put(60,-40){\line(1,2){10}}
\put(27,-40){\mbox{$k$}}
\put(48,-40){\mbox{$l$}}
\put(44,-62){\mbox{$m$}}
\put(120,50){\vector(1,0){50}}
\put(140,55){\mbox{$S_{0j}$}}
\put(270,20){\vector(0,-1){30}}
\put(275,3){\mbox{$S_{0k}\otimes S_{0l}$}}
\put(170,-50){\vector(-1,0){50}}
\put(140,-45){\mbox{$S_{jm}$}}
\end{picture}

\noindent The $T$ transformations can be applied in any of the channels, moving back and forth along this chain.

In the construction of the knot, the typical analytical expression begins from (here the representations and orientations are as in the above diagram):
\be
\label{bridge3gen}
\ldots \
T_m\big[r,\bar r\big]^{a}\
S_{jm}\left[\begin{array}{cc}  k & l  \\ \bar{r}_l  & r_r  \end{array}\right]\
T_k\big[r_l, r\big]^{b}\ T_l\big[\bar r,\bar{r}_l\big]^{c}\
S_{0l}\left[\begin{array}{cc}  \bar{r} & \bar{r}_r  \\ j  & r_r  \end{array}\right]
S_{0k}\left[\begin{array}{cc}  r_l & r  \\ \bar{r}_l  & \bar{j}  \end{array}\right]
S_{0j}\left[\begin{array}{cc}  0 & r   \\ 0  & \bar r  \end{array}\right]
\ee
It is read from right to left, and we can add arbitrary many conjugate and direct $S$ transforms of the same type to the left. In the meantime, whenever two points come close (two external line merge), the necessary number of braidings ($T$ transformations) can be applied.

A few comments are in order. First, note that the obvious selection rule dictates that $j=r$, thus actually there is {\it no} sum, involving the {\it arguments} (not just indices) of the $S$-matrices containing $j$. However, such sums can appear after additional applications of $S$. Second, in the examples with three or more pairs of strands there may appear $S$ and $T$ matrices acting on the non-overlapping vector spaces, like the $T_{k}$ and $T_{l}$ above. In such a case one cannot present the chain of transformations~(\ref{bridge3gen}) as a matrix product. It is rather a multi-matrix, or a tensor product.

\section{Genus $g$ pretzel knots}
\label{sec:generalform}

In this section we will derive some general formulae in terms of the $S$ and $T$ operations. One immediate extension of the examples considered in the previous section, given the method of modular matrices described above, is the series of \emph{pretzel} knots: a combination of 2-strand braids, which can be considered as wrapping around the genus $g$ surface:

\begin{figure}[h!]
\centering\leavevmode
\includegraphics[width=\linewidth]{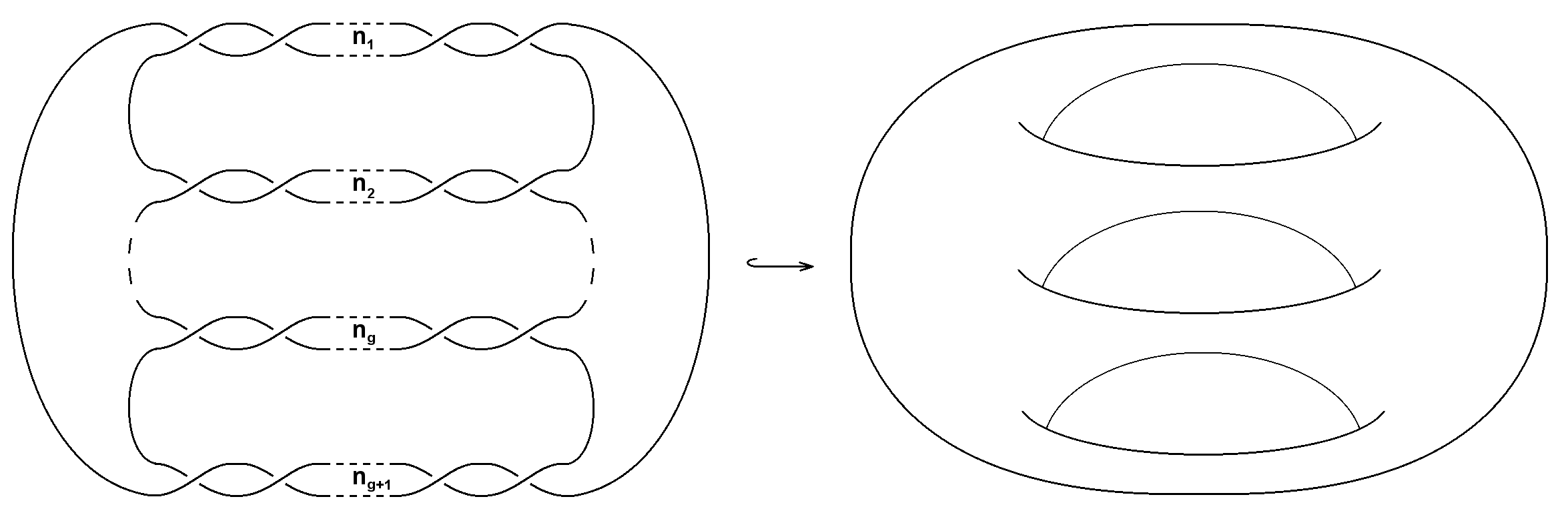}
\caption{Pretzel knot associated with genus $g$ surface}
\label{braidg}
\end{figure}

These knots/links are classified by $g+1$ integers $n_0,\ldots,n_g$, the algebraic lengths of the constituent $2$-strand braids. Orientation of lines does not matter, when one considers the Jones polynomials (not HOMFLY!). Also, these polynomials are defined only for the symmetric representations $[r]$ and, hence, do not change under arbitrary permutations of parameters $n_i$ (though the knot/link itself has at best the cyclic symmetry $n_i\longrightarrow n_{i+1}$). A particular manifestation of this enhanced symmetry has been recently noted in \cite{ensym}.

This family encodes almost all what is currently known about explicit colored knot polynomials in arbitrary symmetric representation beyond torus links: in particular, all the twist and 2-bridge knots are small subsets in the Pretzel family (however, among torus knots with more than two strands, only $[3,4]$ and $[3,5]$ belong to it).

\subsection*{Genus-3 knot  $(n_1,n_2,n_3,n_4)$ and beyond}

 Consider an example using eight-point conformal blocks. Suppose we have ordered the points. The two fundamental conformal blocks necessary for the construction of any braid are shown by the top and the bottom diagrams in the picture below, while the diagram in the middle is the intermediate one, which is related to both of them through a simultaneous application of the $S$-operation in the odd $12$, $34$ etc. channels for the top diagram, or even $23$, $45$ etc. for the bottom one.

\begin{picture}(300,300)(-150,-240)
\put(0,0){\line(1,0){120}}
\put(0,0){\vector(-1,2){10}}
\put(0,0){\vector(-1,-2){10}}
\put(130,20){\vector(-1,-2){5}} \put(125,10){\line(-1,-2){5}}
\put(130,-20){\vector(-1,2){5}} \put(125,-10){\line(-1,2){5}}
\put(16,-10){\mbox{$j_1$}}
\put(59,-10){\mbox{$l$}}
\put(96,-10){\mbox{$j_2$}}
\put(40,0){\line(0,1){20}}
\put(80,0){\line(0,1){20}}
\put(80,20){\vector(1,2){10}}
\put(80,20){\vector(-1,2){10}}
\put(30,40){\vector(1,-2){5}}\put(35,30){\line(1,-2){5}}
\put(50,40){\vector(-1,-2){5}}\put(45,30){\line(-1,-2){5}}
\put(43,12){\mbox{$k$}}
\put(83,12){\mbox{$s$}}
\put(0,-90){\line(1,0){120}}
\put(20,-70){\vector(0,-1){20}}
\put(0,-90){\vector(-1,2){10}}
\put(0,-90){\vector(-1,-2){10}}
\put(50,-70){\vector(0,-1){20}}
\put(70,-90){\vector(0,1){20}}
\put(100,-90){\vector(0,1){20}}
\put(130,-70){\vector(-1,-2){5}} \put(125,-80){\line(-1,-2){5}}
\put(130,-110){\vector(-1,2){5}} \put(125,-100){\line(-1,2){5}}
\put(6,-100){\mbox{$j_1$}}
\put(59,-100){\mbox{$l$}}
\put(106,-100){\mbox{$j_2$}}
\put(32,-84){\mbox{$r$}}
\put(82,-84){\mbox{$r$}}
\put(22,-90){\vector(-1,0){2}}
\put(72,-90){\vector(-1,0){2}}
\put(-18,-110){\mbox{$1$}}
\put(-18,-65){\mbox{$2$}}
\put(19,-65){\mbox{$3$}}
\put(48,-65){\mbox{$4$}}
\put(68,-65){\mbox{$5$}}
\put(98,-65){\mbox{$6$}}
\put(135,-65){\mbox{$7$}}
\put(135,-110){\mbox{$8$}}
\put(0,-200){\line(1,0){120}}
\put(0,-200){\line(0,1){20}}
\put(60,-200){\line(0,1){20}}
\put(120,-200){\line(0,1){20}}
\put(0,-180){\vector(-1,2){10}}
\put(10,-160){\vector(-1,-2){5}}\put(5,-170){\line(-1,-2){5}}
\put(60,-180){\vector(1,2){10}}
\put(50,-160){\vector(1,-2){5}}\put(55,-170){\line(1,-2){5}}
\put(120,-180){\vector(-1,2){10}}
\put(130,-160){\vector(-1,-2){5}}\put(125,-170){\line(-1,-2){5}}
\put(0,-200){\line(0,1){20}}
\put(0,-200){\vector(-1,-2){10}}
\put(130,-220){\vector(-1,2){5}}\put(125,-210){\line(-1,2){5}}
\put(24,-195){\mbox{$r$}}
\put(90,-195){\mbox{$r$}}
\put(2,-200){\vector(-1,0){2}}
\put(62,-200){\vector(-1,0){2}}
\put(3,-188){\mbox{$i_1$}}
\put(63,-188){\mbox{$i$}}
\put(123,-188){\mbox{$i_2$}}
\put(-6,-227){\mbox{$r$}}
\put(124,-227){\mbox{$\bar r$}}
\put(-13,-155){\mbox{$r$}}
\put(9,-155){\mbox{$\bar r$}}
\put(47,-155){\mbox{$\bar r$}}
\put(69,-155){\mbox{$r$}}
\put(107,-155){\mbox{$r$}}
\put(129,-155){\mbox{$\bar r$}}
\end{picture}

\noindent Indeed, the braiding $T$-operation will be trivial in the odd channel of the top diagram, or in the even channel of the bottom diagram. The knot of interest can then be obtained by an appropriate choice of the initial and final conformal blocks.

The two immediate initial and final states are obtained by selecting the top, or the bottom conformal block and assigning the appropriate internal lines the trivial representations. Specifically, setting all the internal lines to be trivial reps in the top diagram gives an appropriate initial and final states for the plat, or quasi-plat presentations, i.e. the TQFT state $|\cup\cup\cup\cup\,\rangle$. One naive family of knots is obtained by applying an arbitrary number of braiding operations in every odd channel of this state, closing the braid again with the $|\cup\cup\cup\cup\,\rangle$ final state. Another family is obtained by taking the bottom conformal block with $i=i_1=i_2=0$ as the initial and final states and applying an arbitrary number of braidings in every even channel. Notice that for both families the initial states are not the natural eigenbases for the $T$-operations.

The second is apparently a larger family of knots and completely includes the first family in it. We will refer to this family as the pretzel knots on genus three surface, or more generally knots on genus $g$ surface in the case of $2g+2$ external points (see figure~\ref{braidg}). All other knots can be obtained by selecting different initial and final states.

Let us now write general expressions for the invariants of the pretzel knot on genus $g$ surface with an associated representation $r$ of $SU_q(N)$ in terms of the $S$ and $T$ matrices. For the $g=3$ case illustrated by the above picture one writes
\be
\sum_{j_1,j_2,k\in r\otimes r}\,\sum_{s\in \bar r\otimes \bar r}
\ \ \sum_{l\in r\otimes \bar r}
S_{j_20}
\left(\begin{array}{cc} r & \bar r \\ r & \bar r\end{array}\right)
S_{j_10}
\left(\begin{array}{cc} r & \bar r \\ r & \bar r\end{array}\right)
S_{l0}\left(\begin{array}{cc} \bar r &  r \\  r & \bar r \end{array}\right)\
S_{sr}\left(\begin{array}{cc} l & r \\ j_{2} & r\end{array}\right) \
 S_{kr}\left(\begin{array}{cc} j_1 & \bar r \\ l & \bar r \end{array}\right)\nn\\
T_{j_1}^{n_0}   T_{k}^{n_1} T_{s}^{n_2}   T_{j_{2}}^{n_3}\ \
S_{rs}\left(\begin{array}{cc}r &  r\\ l & j_{2}\end{array}\right) \
  S_{rk}\left(\begin{array}{cc}\bar r & \bar r \\ j_1 & l\end{array}\right)\
S_{0l}\left(\begin{array}{cc} r & \bar r \\ \bar r & r\end{array}\right)\
S_{0j_1}
\left(\begin{array}{cc}r & r \\ \bar r & \bar r\end{array}\right)
S_{0j_2}
\left(\begin{array}{cc}r & r \\ \bar r & \bar r\end{array}\right) =
\nn
\ee
\be
= \sum_{l\in r \otimes \bar r}  S_{0l}\left(\begin{array}{cc} r & \bar r \\ \bar r &  r\end{array}\right)\
S_{l0}\left(\begin{array}{cc} \bar r & r \\ r & \bar r\end{array}\right) \nn \\
\sum_{j_1,j_2\in r\otimes r}
S_{0j_1}\left(\begin{array}{cc}r & r \\ \bar r & \bar r\end{array}\right)\
S_{j_10}\left(\begin{array}{cc} r & \bar r \\ r & \bar r\end{array}\right)
S_{0j_2}\left(\begin{array}{cc}r & r \\ \bar r & \bar r\end{array}\right)
S_{j_20}\left(\begin{array}{cc} r & \bar r \\ r & \bar r\end{array}\right) \
T_{j_1}^{n_0}(r,r)\  T_{j_{2}}^{n_3}(\bar r,\bar r) \ \  \nn \\
\sum_{k\in \bar r\otimes \bar r}
S_{rk}\left(\begin{array}{cc}\bar r & \bar r \\ j_1 & l\end{array}\right)\
T_{k}^{n_1}(\bar r,\bar r) \
S_{kr}\left(\begin{array}{cc}j_1 & \bar r \\ l & \bar r\end{array}\right)
\sum_{s\in  r\otimes  r}
S_{rs}\left(\begin{array}{cc}r &  r\\ l & j_{2}\end{array}\right)\
T_{s}^{n_2}(r,r) \
S_{sr}\left(\begin{array}{cc}l & r \\ j_{2} & r\end{array}\right)
\ee
where we set $i=i_1=i_2=0$.
We remind that we construct the knot, first, applying the $S$-transformations to change from the initial state to a convenient basis of conformal blocks, then applying the $T$-operations and finally contracting the result with the final state.

This can be extended to the general $g$ case. For the number of braidings in every odd channel defined by the vector $(n_0,n_1,\ldots,n_{g})$,
\be
\sum_{j_1,j_2,\ldots,\in r\otimes r\atop l_1,l_2\ldots \in r\otimes \bar r}
 \prod_{m=1}^{\left[\frac{g}{2}\right]}
S_{0l_m}\left(\begin{array}{cc} r & \bar r \\ \bar r &  r\end{array}\right)
S_{l_m0}\left(\begin{array}{cc} \bar r &  r \\ r & \bar r\end{array}\right)
\prod_{i=1}^{\left[\frac{g+1}{2}\right]}
S_{0j_i}\left(\begin{array}{cc}r & r \\ \bar r & \bar r\end{array}\right)
S_{j_i0}\left(\begin{array}{cc}r & \bar r \\ r & \bar r\end{array}\right) T_{j_1}^{n_0}(r,r) T_{\rm last}^{n_g}\times
 \nn \\
\sum_{k_1,k_2\ldots \in \bar r\otimes \bar r\atop s_1,s_2\ldots \in r\otimes r}
\prod_{q=1}^{\left[\frac{g}{2}\right]}
S_{rk_{q}}\left(\begin{array}{cc} \bar r & \bar r \\ j_q & l_{q}\end{array}\right)
T_{k_q}^{n_{2q-1}}(\bar r,\bar r)
S_{k_{q}r}\left(\begin{array}{cc} j_q & \bar r \\ l_q & \bar r \end{array}\right)
\prod_{p=1}^{\left[\frac{g-1}{2}\right]}
 S_{rs_{p}}\left(\begin{array}{cc} r & r \\ l_p & j_{p+1} \end{array}\right)
 T_{s_p}^{n_{2p}}(r,r)
S_{s_{p}r}\left(\begin{array}{cc} l_p & r \\ j_{p+1} & r\end{array}\right)
\label{genScomb}
\ee
where the square brackets $[\cdot]$ in the upper limits of the product denote the integer part of a fraction and $T_{\rm last}$ is the braiding in the last channel, i.e. $T_{\rm last}=T_{j_\frac{g+1}{2}}(\bar r,\bar r)$, if $g$ is odd, and $T_{\rm last}=T_{l_\frac{g}{2}}(r,r)$, if $g$ is even.

In the case of $SU_q(2)$ $r=\bar r$. Equation~(\ref{genScomb}) can be further simplified as follows
\be
\label{genScombSU2}
\begin{split}
\langle n_0,\ldots,n_g \rangle=\sum\limits_{j_l, \; l=1,\ldots g}
\left(\prod\limits_{i=1}^{g}
S_{0j_i}\left[\begin{array}{cc}  r & r\\ r & r\\ \end{array}\right]
S_{j_i0}\left[\begin{array}{cc}  r & r\\ r & r\\ \end{array}\right]
\right)\ T_{j_1}^{n_0}(r,r)\, T_{j_g}^{n_g}(r,r)\times\\
\sum\limits_{k_m, \; m=1,\ldots g-1}\left( \prod\limits_{p=1}^{g-1}
S_{rk_p}\left[\begin{array}{cc} r & r \\ j_{p} & j_{p+1}\\ \end{array}\right]
T_{k_p}^{n_{p}}(r,r)
S_{k_pr}\left[\begin{array}{cc} j_{p} &  r\\ j_{p+1} & r \\ \end{array}\right]\right)
\end{split}
\ee
where the indices $j_l$ and $k_m$ run over the set parameterized by irreps that appear in the tensor product $r\otimes r$. Indeed, as $r=\bar r$, there is no difference between $j$ and $l$, or $k$ and $s$ types of indices.

In the fundamental representation of $SU_q(2)$, the $S$ and $T$-matrices are given by equation~(\ref{ST2}). The indices in the above formulae will run over the set $([2],0)=[1]\otimes [1]$. The other necessary Racah matrices can be computed via the methods outlined in section~\ref{sec:ST}, or taken explicitly from~\cite{Racah,AG}. Let us quote few examples that follow from equation~(\ref{genScombSU2}):

\paragraph{genus 1:}
     \be
     \label{genus1ex0}
     \langle n_0,n_1\rangle\equiv {J_{1}^{(n_0,n_1)}\over [2]^2}=\frac{1}{[2]^2}\left(\lambda_0^{n_0+n_1}+[3] \lambda_1^{n_0+n_1} \right)
     \ee
This indeed agrees with~(\ref{2straN2}) and~(\ref{2straN2kn}).
\paragraph{genus 2:}
     \be
        \langle n_0,n_1,n_2\rangle\equiv {J_{1}^{(n_0,n_1,n_2)}\over [2]^3}=\frac{1}{[2]^4}\left(\lambda_0^{n_0+n_1+n_2}+[3]\left(\lambda_0^{n_0} \lambda_1^{n_1+n_2}+{\rm permutations}\right)+\left([5]+1\right)\lambda_1^{n_0+n_1+n_2}\right)
   \label{gen2f}  \ee
where ``permutations'' stand for the combination that symmetrizes the expression in the square brackets with respect to permutations of $n_i$.

\paragraph{genus 3:}
    \be
    \label{genus3ex0}
         \langle n_0,n_1,n_2,n_3\rangle\equiv {J_{1}^{(n_0,n_1,n_2,n_3)}\over [2]^4} =\frac{1}{[2]^6}\left(\lambda_0^{n_0+n_1+n_2+n_3}+[3]\left(\lambda_0^{n_0+n_1} \lambda_1^{n_2+n_3}+{\rm permutations}\right)+\right. \nn \\
         +\left. \left([5]+1\right)\left(\lambda_0^{n_0}\lambda_1^{n_1+n_2+n_3}+{\rm permutations}\right) +([7]+[5]+3[3])\lambda_1^{n_0+n_1+n_2+n_3}\right)
    \ee

From the examples one can already notice that the invariants have a certain structure. One can attempt to generalize this formulae for an arbitrary genus, as well as arbitrary representation of $SU_q(2)$. This will be done in the next section.

\section{Jones polynomials for the pretzel knots on genus $g$ surface}
\label{sec:polynomials}

In \cite{Sle} an ambitious suggestion was made to look for a generalization of the Rosso-Jones formula~\cite{RJ} from torus to the pretzel knots on genus $g$ surface. With an unexpected success, this program was approached by the evolution method of~\cite{DMMSS} and \cite{evo}. The first version of ref.~\cite{Sle} (which consisted of first two sections and Appendix A of the present version) provides impressive results, clearly implying a deep structure in knot polynomials, indeed comparable with the Rosso-Jones formula. However, computations there originally turned out to be too difficult to reveal this structure completely, and only using the results for the Jones polynomials of this paper we managed to construct remarkable general formulas of the present version of ref.~\cite{Sle}.

In~\cite{GMMMS} the authors of this work proposed a general formula for the colored Jones and HOMFLY polynomials of the pretzel knots on genus $g$ surface. The formulae appeared as a summary of a large number of calculated examples via the $S$-matrix method reviewed above. In the present paper we describe the results of these calculations and explain the steps that lead to the proposed formula in the case of Jones polynomials. The extension to HOMFLY polynomials was done in~\cite{Sle}.

The purpose of this section is to reveal completely the structure of the polynomials of the arbitrary pretzel knots, and once again, to demonstrate the power of the $S$-matrix technique, though it is only in combination with other methods that it leads to real success.

\subsection*{The steps of the calculation}

Evaluation of every particular colored Jones along the lines of section~\ref{sec:generalform} is straightforward. However, since we are hunting for a general structure, the choice of things to compute should be more systematic.
Actually we go from one observation to another, also mixing them with the insights from the HOMFLY studies in the first version of \cite{Sle}. In result the sequence is as follows.

\paragraph{1.} The pretzel knot on genus $g$ surface is labeled by $g+1$ integers $n_1,\ldots,n_{g+1}$. There is an additional dependence on the mutual orientation of the strands in each constituent braid, but Jones polynomials do not feel it:\footnote{
Up to framing factors, since the claim is based on the group argument that the irrep and its conjugate coincide in $SU_q(2)$, and the group theory corresponds to the vertical framing. The vertical framing corresponds to the choice of (\ref{fr}) with $C_R=\sum_{i,j\in R}(j-i)$.} the quantity of interest is just the (unreduced) Jones polynomials
\be
J_{r}^{(n_1\ldots n_{g+1})}(q) =  H_{[r]}^{(n_1\ldots n_{g+1})}(q,A=q^2)\,
\ee
Again, we identify the number of boxes $r$ in the representation $[r]$ of $SU_q(2)$ with the ``spin'', although the actual spin is $r/2$. As explained in \cite{Sle}, the naive cyclic symmetry $n_i\longrightarrow n_{i+1}$ of the knot diagram, is actually enhanced to the full permutation symmetry: this is because permutation  $n_i\leftrightarrow n_{i+1}$ replaces the knot/link for its mutant cousin, undistinguishable by knot polynomials in symmetric representations, and thus by arbitrary colored Jones polynomials.

\paragraph{2.} The evolution method implies that the $n_i$ dependence is given by
\be
J_{r}^{(n_1\ldots n_{g+1})} =
\sum_{m_1,\ldots,m_{g+1}=0}^r \tilde C_{m_1,\ldots,m_{g+1}}
\lambda_{m_1}^{n_1}\ldots \lambda_{m_{g+1}}^{n_{g+1}}\,
\ee
where $\lambda_m$ are made from the eigenvalues $\varkappa_{[r+m,r-m]}=\varkappa_{[r+m,r-m]}([2])=C_m$
of the cut-and-join operator $\hat W([2])$ (the simplest from the family of \cite{MMN},
$\hat W(\Delta) \chi_R = \varphi_R(\Delta)\chi_R$) on the representation of $GL(N)$ given by the Young diagram $[r+m,r-m]$ (which corresponds to the spin $2m$ representation of SU(2)):
\be
\lambda_m = (-)^mq^{m(m+1)}\,
\ee
up to an arbitrary framing factor depending on $r$.

\paragraph{3.} Enhanced symmetry implies that the coefficient tensors $\tilde{C}_{\ldots}$ are totally symmetric,
which allows one to switch to a dual parametrization,
where the sum is over multiplicities $j_m$ of the eigenvalues $\lambda_{r-m+1}$:
\be
\label{coefparam}
J_{r}^{(n_1\ldots n_{g+1})} = \frac{1}{[r+1]^{g-1}}
\sum\limits_{j_1,\ldots,j_r=0}^{j_1+\ldots+j_r\leq g+1}
C_{j_1,\ldots,j_r}\cdot\Big(
\lambda_r^{n_1+\ldots +n_{j_1}}\lambda_{r-1}^{n_{j_1+1}+\ldots+n_{j_1+j_2}}\ldots
\lambda_0^{n_{j_1+\ldots +j_r+1}+\ldots+n_{g+1}} +\nn \\
+ \text{permutations of\ }n'{\rm s}\Big)\,
\ee
We agree to sum only over ``active" permutations, i.e. do not permute parameters $n_i$ within the sets $\{n_1,\ldots,n_{j_1}\}$, $\{n_{j_1+1},\ldots,n_{j_1+j_2}\}$, $\ldots$: e.g. the term $C_{r0\ldots 0}\lambda_r^{n_1+\ldots+n_{g+1}}$ is {\it not} multiplied by $(g+1)!$.

\paragraph{4.} Already the first explicit examples~(\ref{genus1ex0})-(\ref{genus3ex0}) indicate that the coefficients $C_{j_1,\ldots,j_{r}}$ are independent of $g$, i.e. the dependence on $\lambda_0$ in genus $g$ is fully determined by the answer in genus $g-1$. All non-zero coefficients, which appear in genus $g-1$ are imported to genus $g$ as coefficients of some $\lambda_0$ term.

\paragraph{5.} Whenever any of the $n_i$ vanishes, the knot/link becomes a composite. This means that the knot polynomial factorizes into a product -- in fact into a product of $g$ elementary $2$-strand polynomials
\be
J_{r}^{(n)} = \sum_{m=0}^r [2m+1]\, \lambda_m^n\,
\ee
This means that the difference
\be
J_{r}^{(n_1\ldots n_{g+1})} - \prod_{i=1}^{g+1} J_{r}^{(n_i)}\,
\ee
vanishes whenever $n_i=0$ for any $i$, and it is instructive to see what makes this {\it compositeness constraint} satisfied in the final answers.

\paragraph{6.} In the fundamental representation $r=1$, the difference is yet another product \cite{Sle}:
\be
J_{1}^{(n_1\ldots n_{g+1})} - \prod_{i=1}^{g+1} J_{1}^{(n_i)} = \frac{[3]}{[2]^{g+1}} \prod_{i=1}^{g+1}\Big( -\lambda_1^{n_i}+\lambda_0^{n_i}\Big)\,
\ee
which obviously vanishes when any $n_i=0$. As to the formula for the Jones polynomial itself, this implies that
\be
C_i = {1\over [2]^2}\left([3]^i\ +\ (-)^i [3]\right)
\ee
is a sum of two powers, and
\be
J_{1}^{(n_1\ldots n_{g+1})} =
\frac{1}{[2]^{g-1}}\,\sum_i C_i \cdot \left\{\lambda_{1}^{n_1+\ldots+n_i}\lambda_{0}^{n_{i+1}+\ldots+n_{g+1}}
\ + \ \text{permutations of}\ n'{\rm s}  \right\}
= \nn \\
=\frac{1}{[2]^{g+1}}
\left\{\ \prod_{i=1}^{g+1}\Big([3]\lambda_{1}^{n_i} + \lambda_{0}^{n_i}\Big)
\ + \ [3]\prod_{i=1}^{g+1}\Big(-\lambda_{1}^{n_i} + \lambda_{0}^{n_i}\Big)\right\}
\label{J1}
\ee

\paragraph{7.} Similarly, in the representation $[2]$
\be
J_{2}^{(n_1\ldots n_{g+1})} = \frac{1}{[3]^{g-1}} \sum\limits_{{i,j=0}}^{i+j\leq g+1}
C_{ij}\cdot\left\{\lambda_{2}^{n_1+\ldots+n_i}\lambda_{1}^{n_{i+1}+\ldots+n_{i+j}}
\lambda_{0}^{n_{i+j+1}+\ldots +n_{g+1}}
\ + \ \text{permutations of}\ n'{\rm s}\right\}\,
\ee
with
\be
 C_{ij} = \frac{1}{[3]^2}\left\{ [5]^i[3]^{j}
 + \left(\frac{[2]}{[4]}\right)^{i+j}
 \left((-)^j[3]^j[5]+(-)^i [3][5]^i\left(\frac{[6]}{[2]}\right)^j\right)\right\}
\ee
i.e.
\be
J_{2}^{(n_1\ldots n_{g+1})} = \cfrac{1}{[3]^{g+1}} \left\{\ \prod_{i=1}^{g+1}\Big([5]\lambda_{2}^{n_i} +
[3]\lambda_{1}^{n_i} + \lambda_{0}^{n_i}\Big)
\ + \right.\nn \\ \nn \\ \left.
\ + \ [3]\prod_{i=1}^{g+1}\left(-\cfrac{[2][5]}{[4]}\lambda_{2}^{n_i} + \cfrac{[6]}{[4]}\lambda_{1}^{n_i} + \lambda_{0}^{n_i}\right)
\ + \ [5]\prod_{i=1}^{g+1}\left(\cfrac{[2]}{[4]}\lambda_{2}^{n_i} - \cfrac{[2][3]}{[4]}\lambda_{1}^{n_i} + \lambda_{0}^{n_i}\right)\right\}
\label{J2}
\ee
is a sum of three products. The first one involves the ordinary 2-strand polynomial
\be
J_2^{(n)} = \frac{1}{[3]^2}\left([5]\lambda_2^n + [3]\lambda_1^n+\lambda_0^n\right)
\ee
while the two other involve the two ``satellite" polynomials
\be
J_{2,1}^{(n)} = \frac{1}{[3]^2}\left(-\cfrac{[2][5]}{[4]}\,\lambda_{2}^{n}\ + \ \cfrac{[6]}{[4]}\,\lambda_{1}^{n}\ +\ \lambda_{0}^{n}\right)
\ee
and
\be
J_{2,2}^{(n)}=\frac{1}{[3]^2}\left(\cfrac{[2]}{[4]}\,\lambda_{2}^{n}\ - \ \cfrac{[2][3]}{[4]}\,\lambda_{1}^{n} \ +\ \lambda_{0}^{n}\right)
\ee
The compositeness constraint is satisfied, because the sums of the coefficients of each satellite is zero:
\be
 -\cfrac{[2][5]}{[4]} \   +\ \cfrac{[6]}{[4]} \  +\  1 \ =\  0\,, \nn \\
\cfrac{[2]}{[4]} \  -\ \cfrac{[2][3]}{[4]}\ +\ 1\ =\  0\,
\label{ortho2}
\ee
Moreover, the two satellites are ``orthogonal" in an appropriate metric:
\be
-\cfrac{[2][5]}{[4]}\cdot[5]^{-1}\cdot  \cfrac{[2]}{[4]}
\ \ +\ \  \cfrac{[6]}{[4]}\cdot[3]^{-1} \cfrac{[2]}{[4]}
\ \ +\ \  1 = 0\,
\ee
Consequently, equations~(\ref{ortho2}) can be interpreted as the orthogonality of the satellites to the parent 2-strand Jones.

\paragraph{8.} We are now ready to make the general conjecture: the genus-$g$ Jones polynomials in the representation $[r]$ is a sum of $r+1$ products
\be
\boxed{
J_r^{(n_1\ldots n_{g+1})}  = \frac{1}{[r+1]^{g+1}} \sum_{k=0}^r \ [2k+1]\cdot\prod_{i=1}^{g+1}
\left(\sum_{m=0}^r  a_{km} \lambda_m^{n_i}\right)
}
\label{Jgrfirst}
\ee
or
\be
C_{j_1\ldots j_r} =
\frac{1}{[r+1]^2}\,\sum_{k=0}^r \  [2k+1]\cdot a_{k1}^{j_1}\cdot a_{k2}^{j_2} \cdot\ldots\cdot a_{kr}^{j_r}\,
\ee
The first product is made out of the 2-strand Jones, i.e.
\be
a_{0m} = [2m+1]\,
\ee
The $r$ ``satellite" polynomials satisfy orthogonality conditions
\be
\boxed{
\frac{1}{[r+1]^2}\,\sum_{m=0}^{r} \frac{a_{km}a_{k'm}}{[2m+1]} = \frac{\delta_{k.k'}}{[2k+1]}
}
\ \ \ \ \forall \ k,k'=0,\ldots, r\,
\ee
and their $k'=0$ component guarantees the validity of the compositeness constraint.

\paragraph{9.} In fact, there are also dual orthogonality relations:
\be
\boxed{
\frac{1}{[r+1]^2}\,\sum_{k=0}^{r} [2k+1]\cdot {a_{km}a_{km'}}  = [2m+1]\,\delta_{m.m'}
}
\ \ \ \ \ \forall \ m,m'=0,\ldots,r\,
\ee
which follow from the genus-one result, which is again the elementary $2$-strand polynomial $J_r^{(n_1,n_2)}$, only depending on the sum $n_1+n_2$, and the dual ``boundary conditions"
\be
a_{k0} = 1\,
\label{Jgrlast}
\ee

\paragraph{10.} Altogether this makes ${\cal A}_r=\{a_{km}\}$ a quasi-unitary matrix (quasi -- because of non-trivial ``metric" $\{[2m+1]\}$),
rotating the powers of eigenvalues $\lambda_m$ into two-strand Jones polynomials and their satellites. In the first three representations, one finds
\be
{\cal A}_1 = \left(\begin{array}{cc} 1 & [3] \\ 1 & -1 \end{array}\right), \nn \\
\nn \\ \nn \\
{\cal A}_2 = \left(\begin{array}{ccc} 1 & [3] & [5] \\ \\
1 & \frac{[6]}{[4]} & -\frac{[2][5]}{[4]} \\ \\
1& -\frac{[2][3]}{[4]} & \frac{[2]}{[4]}  \end{array}\right), \nn \\
\nn \\ \nn \\
{\cal A}_3 = \left(\begin{array}{cccc} 1 & [3] & [5] & [7] \\ \\
1 & \frac{[2][6]-1}{[5]}  & [5]-[2]^2 &  -\frac{[3][7]}{[5]}\\ \\
1& \frac{[7]-[2]^2}{[5]}  & -\frac{[2]\,\big([2][6]-[3]\big)}{[6]}
&  \frac{[2][3][7]}{[5][6]} \\ \\
1 & -\frac{[3]^2}{[5]} & \frac{[2][3]}{[6]}  &  -\frac{[2][3]}{[5][6]}
\end{array}\right), \nn \\ \nn\\
\ldots
\ee
What should these rotation matrices be for general $r$? It is difficult to miss the similarity between ${\cal A}_1$ and the $SU(2)$ $S$-matrix given by equation (\ref{ST2}). Moreover, from these first examples, it looks like there is a general relation between ${\cal A}_r$ and the Racah matrices:
\be
\label{Arel}
\boxed{
a_{km} =\  [r+1]\sqrt{\frac{[2m+1]}{[2k+1]}}\cdot
S_{km} \left(\begin{array}{cc}r &r\\ r &r \end{array}\right)
}
\ee
where $S_{km} \left(\begin{array}{cc}r_2 &r_3\\ r_1 &r_4 \end{array}\right)$ denotes the Racah matrix for $SU_q(2)$ \cite{Racah,AG}.
It is this formula that we claim as our final answer. It would be interesting though to understand, why it has such a simple form. The simplicity hints the existence of a more straightforward, perhaps one-line derivation of the same result. This is, however, beyond the scope of the present paper.

\paragraph{11.} Originally we calculated directly many entries in representations $r\leq 5$, and this allowed to discover the above structure and formulate the orthogonality relations. However, once they are known, the problem simplifies a lot. The thing is that the complexity of formulas for $a_{km}(q|r)$ increases from the perimeter of the matrix ${\cal A}$ to its center. One can inductively find generic-$r$ expressions at the perimeter
 and use orthogonality relations to calculate the entries closer to the center, thus getting new material for induction at deeper and deeper layers.
Tedious $S$-matrix calculations are no longer needed, only to check the conjectured formulas.

In the Appendix, we collect some interesting intermediate formulas, mentioned in the above line of reasoning and provide a few more entries of matrices ${\cal A}_r$. Spectacularly, all this structure survives lifting from the Jones to
HOMFLY polynomials \cite{Sle}.

\section{Conclusions}

In this paper we have applied the $S$-matrix method to the derivation of the colored Jones polynomials of a large series of knots. Specifically, we have studied the family of pretzel knots and links on genus $g$ surface, which can be viewed as a natural extension of the 2-strand torus case. Summarizing the large number of examples studied, we have explained the origin of the general formulae~(\ref{Jgrfirst}) and~(\ref{Arel}), \emph{\`a la} Rosso-Jones, for this family of knots, which appeared in the companion paper~\cite{GMMMS}.

The method based on the use of the ($SL(2,\mathbb{Z})$) modular matrices $S$ and $T$, to which we referred above as the $S$-matrix method, proved to be a very efficient tool in the computation of knot polynomials. Besides allowing to construct and implement algorithms to compute the invariants (sections~\ref{sec:examples} and~\ref{sec:generalform}), it is self-contained since it provides a unique framework for both finding the elementary building blocks, such as the $S$ and $T$ matrices themselves, and using them in the construction. As we stress in section~\ref{sec:STmatrices} rational $W_N$ models is a natural playground to build the matrix representations of the $S$ and $T$ operators.

The final formula~(\ref{Jgrfirst}) relates the invariants at given representation $r$ of $SU(2)$ to the components of a single Racah ($S$) matrix in a remarkably simple way~(\ref{Arel}). The simplicity of the relation indicates that a much simpler derivation, and perhaps a proof of the general formula is somewhere around the corner. In the current form the proposed relation may be regarded as a consistency check for the known Racah matrices, or perhaps, even a new way to derive them.

We remind that in the proposed derivation an important role was played by the parallel study~\cite{Sle}, where the Jones polynomials were uplifted to the form of HOMFLY. This generalization opens many new perspectives. The new interesting possibilities may include a new kind of decomposition of knot polynomials into the Racah rotated elementary Jones (HOMFLY), emerging connection to the (modular transformations of the) \emph{toric} conformal blocks, a kind of monopole/brane duality and so on. Clearly one may also try to generalize the above family to new series of knots. It would be interesting to see whether the simplicity of the relation described in this paper could be maintained. The results of these excursions will be reported elsewhere.

\section*{Acknowledgements}

We would like to thank the hospitality of the International Institute of Physics -- UFRN and the workshop ``Group Theory and Knots'', where a part of this work was done.

Our work is partly supported by grants NSh-1500.2014.2, by RFBR grants 13-02-00457 (D.G. \& A.Mir.), 13-02-00478 (A.Mor.),  14-02-00627 (D.M.), by the joint grants 13-02-91371-ST-a and 15-51-52031-NSC-a (A.Mir.\& A.Mor.), by 14-01-92691-Ind-a (D.M., A.Mir. \& A.Mor.),
15-52-50041-YaF (D.G., A.Mir. \& A.Mor.), and by the young-scientist grant 14-01-31395-young-a (D.G.). Also we are partly supported by the Brazilian Ministry of Science, Technology and Innovation through the National Counsel of Scientific and Technological Development (D.M. \& A.Mor.) and by DOE grants SC0010008, ARRA-SC0003883, DE-SC0007897 (D.G.).

\newpage

\section*{Appendix}

In this Appendix we provide a number of examples to demonstrate the key observations made in the section 5. An the end of section~\ref{sec:generalform}, we have spelled the results for the fundamental representation at genera $g=1,2,3$. We provide some results for other representations as well. Let us start from writing the formula for the $r=1$ and $g=4$ Jones polynomial,
\begin{multline}
 J_{1}^{(n1,\ldots,n_5)} = \frac{1}{[2]^3}\Bigg(\lambda_0^{\sum_j n_j} +[3]\left(\lambda_0^{n_1+n_2+n_3}\lambda_1^{n_4+n_5}+\text{perm.}\right)
 +  ([5]+1)\left(\lambda_0^{n_1+n_2}\lambda_1^{n_3+n_4+n_5}+\text{perm.}\right) \\
 + \left([7]+[5]+3[3]\right)\left(\lambda_0^{n_1}\lambda_1^{n_2+n_3+n_4+n_5}+\text{perm.}\right)
 +  \left(2\,\frac{[3]^2[4]}{[2]}+\frac{[3][4]^3}{[2]^3}\right)\lambda_1^{\sum_jn_j}\Bigg).
 \end{multline}
As observation~{\bf 4} states, the coefficients of the $\lambda_0$ containing terms all appear in examples~(\ref{genus1ex0})-(\ref{genus3ex0}).

In the representation $r=2$, one obtains

\paragraph{genus 1:}
\be
\label{g1r2}
J^{(n_1,n_2)}_2 = \lambda_0^{n_1+n_2}+
[3]\lambda_{1}^{n_1+n_2}+[5]\lambda_2^{n_1+n_2}
\ee

\paragraph{genus 2:}
\begin{multline}
J_2^{(n_1,n_2,n_3)}  =  \frac{1}{[3]}\Bigg(\lambda_0^{n_1+n_2+n_3}+[3]\left(\lambda_0^{n_1}\lambda_1^{n_2+n_3} + \text{permutations}\right) +
\\  +\frac{[3]^2}{[2]^2[4]^2}\left(1-2[5]+[5]^2\right)\lambda_1^{n_1+n_2+n_3}+[5]\left(\lambda_0^{n_1}\lambda_2^{n_2+n_3}+\text{permutations}\right) + [3][5]\,\frac{[2]^2}{[4]^2}\left(\lambda_1^{n_2+n_3}\lambda_2^{n_1}+\text{permutations}\right) \\
+ \frac{[2][3][5][6]}{[4]^2}\left(\lambda_1^{n_1}\lambda_2^{n_2+n_3}+\text{permutations}\right) + \frac{[2][5][6][7]}{[3][4]^2}\lambda_2^{n_1+n_2+n_3}\Bigg)
\end{multline}
\paragraph{genus 3:}
\begin{multline}
J_2^{(n_1,n_2,n_3,n_4)}  =  \frac{1}{[3]^2}\Bigg(\lambda_0^{n_1+n_2+n_3+n_4}+[3]\left( \lambda_0^{n_1+n_2} \lambda_1^{n_3+n_4} + \text{perm.}\right) +
\\ +\frac{[3]^2([5]-1)^2}{[2]^2[4]^2}\left( \lambda_0^{n_1} \lambda_1^{n_2+n_3+n_4} + \text{perm.}\right) +\left(\frac{[3]^3([5]-1)^4}{[2]^4[4]^4}+[3]^2+\frac{[3]^2[2]^4[5]}{[4]^4}\right)\lambda_1^{n_1+n_2+n_3+n_4}+
\\ + [5]\left( \lambda_0^{n_1+n_2} \lambda_2^{n_3+n_4} + \text{perm.}\right) +\left([3][5]+\frac{[3]^2[5][6]([5]-1)^2}{[2][4]^4}+\frac{[2]^3[5][6][7]}{[4]^4}\right)\left( \lambda_1^{n_1+n_2} \lambda_2^{n_3+n_4} + \lambda_1^{n_3+n_4} \lambda_2^{n_1+n_2}\right)+
\\ + \frac{[2]^2[3][5]}{[4]^2}\left(\lambda_0^{n_1}\lambda_1^{n_2+n_3}\lambda_2^{n_4}+\text{perm.}\right) + \left(\frac{[3]^2[5]([5]-1)^2}{[4]^4}+\frac{[2]^3[3]^2[5][6]}{[4]^4}\right)\left( \lambda_2^{n_1} \lambda_1^{n_2+n_3+n_4} + \text{perm.}\right)+
\\ + \left(\frac{[2]^4[3][5]^2}{[4]^4}+\frac{[2]^2[3]^2[5][6]^2}{[4]^4}\right)\left( \lambda_1^{n_1+n_3} \lambda_2^{n_2+n_4} + \lambda_1^{n_1+n_4} \lambda_2^{n_2+n_3} + \lambda_2^{n_1+n_3} \lambda_1^{n_2+n_4} + \lambda_2^{n_1+n_4} \lambda_1^{n_2+n_3}\right) +
\\ + \frac{[2][3][5][6]}{[4]^2}\left(\lambda_0^{n_1}\lambda_1^{n_2}\lambda_2^{n_3+n_4}+\text{perm.}\right)+
\left(\frac{[2]^3[3][5]^2[6]}{[4]^4}+\frac{[2]^2[5][6]^2[7]}{[4]^4}\right)\left( \lambda_1^{n_1} \lambda_2^{n_2+n_3+n_4} + \text{perm.}\right)+
\\  + \frac{[2][5][6][7]}{[3][4]^2}\left(\lambda_0^{n_1}\lambda_2^{n_2+n_3+n_4}+\text{perm.}\right)+ [5]\left([5]+\frac{[2]^2[3][5][6]^2}{[4]^4}+\frac{[2]^2[6]^2[7]^2}{[3]^2[4]^4}\right)\lambda_2^{n_1+n_2+n_3+n_4}\Bigg).
\end{multline}
Due to complicated coefficients and the way the partitions of the string $(n_1,n_2,n_3,n_4)$ appear, the last result seems to be non-trivial enough to test the general formulae proposed in observations~{\bf 5} and~{\bf 7}.

For $r=3$ we just spell the $g=1,2$ results:
\paragraph{genus 1:}
\be
J^{(n_1,n_2)}_2 = \lambda_0^{n_1+n_2}+
[3]\lambda_{1}^{n_1+n_2}+[5]\lambda_2^{n_1+n_2}+[7]\lambda_3^{n_1+n_2}
\ee
At this step one can be convinced that the his/her guess about the general formula for $g=1$, made after the $g=1$ and $r=2$ example~(\ref{g1r2}), was correct (see observation~{\bf 9}).

\paragraph{genus 2:}
\begin{multline}
J_3^{(n_1,n_2,n_3)}  =  \frac{1}{[4]}\Bigg(\lambda_0^{n_1+n_2+n_3}+[3]\left(\lambda_0^{n_1}\lambda_1^{n_2+n_3} + \text{permutations}\right) +
\\  +\frac{[3]}{[4][5]}\left([2]-2[6]+\frac{[6]^2}{[2]}\right)\lambda_1^{n_1+n_2+n_3}+[5]\left(\lambda_0^{n_1}\lambda_2^{n_2+n_3}+\text{permutations}\right)+
\\+ [7] \left(\lambda_0^{n_1}\lambda_3^{n_2+n_3}+\text{permutations}\right)+
\frac{[3]^2[7][8]}{[5][6]} \left(\lambda_1^{n_1}\lambda_3^{n_2+n_3}+\text{permutations}\right)
\\ + [6]\,\frac{[2]^4}{[4][5]}\left(\lambda_1^{n_2+n_3}\lambda_2^{n_1}+\text{permutations}\right)+ \frac{[2]^2}{[4][6]}\left(1-2[7]+[7]^2\right)\left(\lambda_1^{n_1}\lambda_2^{n_2+n_3}+\text{permutations}\right) +
\\  + \frac{[2][5][7]}{[3]^2[4][6]^2}\left({[2]^6}-2{[2]^3[8]} +{[8]^2}\right)\lambda_2^{n_1+n_2+n_3} +
\frac{[2][3][7][8][9]}{[5][6]^2} \left(\lambda_2^{n_1}\lambda_3^{n_2+n_3}+\text{permutations}\right)
\\ +
\frac{[2]^2[3]^2[7][8]}{[4][6]^2} \left(\lambda_2^{n_2+n_3}\lambda_3^{n_1}+\text{permutations}\right) + \frac{[2][3][7][8][9][10]}{[4][5]^2[6]^2} \lambda_3^{n_1+n_2+n_3}+
\\ + \frac{[2][3]^2[7]}{[5][6]}\left(\lambda_3^{n_3}\lambda_2^{n_1}\lambda_1^{n_2}+\text{permutations}\right)\Bigg).
\end{multline}

At this stage, higher values of $g$ and $r$ seem to be too bulky to present here. One should be convinced that the coefficients are indeed independent of the genus and it makes sense to write the general result in the form of equation~(\ref{coefparam}) of observation~{\bf 3}, where the coefficients depend only on the representation, otherwise universal for all genera. For the representation $r=3$, the following ansatz can be used

\be
C_{ijk} = \frac{1}{[4]^2} \Big([7]^i[5]^j[3]^k
\  + \  [3]\cdot a_{1}^i\cdot b_{1}^j\cdot c_1^k\ +\   [5]\cdot a_{2}^i\cdot b_{2}^j\cdot c_2^k
+  [7]\cdot a_3^i\cdot b_3^j\cdot c_3^k\Big).
\ee
This is in fact a set of equations, which one can solve for the parameters $a_i$, $b_i$ and $c_i$ after selecting a sufficient number of coefficients $C_{ijk}$. Although naively the equations are cubic, an educated choice of the coefficients allows to subsequently reduce the system to the second order equations. Our study of a number of examples with $r=3$ and $i+j+k\leq g+1 = 4$ allowed to find the parameters $a_i$, $b_i$ and $c_i$, that yielded corresponding invariants:
\be
 a_1 = -\frac{[3][7]}{[5]}, \ \ \ \
 & b_1= \frac{[2][12]}{[4][6]}=\frac{q^8+1+q^{-8}}{q^4+1+q^{-4}}=\frac{[3]_4}{[3]_2}
 = q^4-1+q^{-4},
 \ \ \ \  & c_1=\frac{[7]+[5]-1}{[5]},\nn \\
 a_2 = \frac{[2][3][7]}{[5][6]}, \ \ \ \
 & b_2=-\frac{[2][3]}{[6]}\cdot\Big([5]-2\Big)
 =-\frac{[3](q^4+q^2-1+q^{-2}+q^{-4})}{[3]_2}, \ \ \ \
&  c_2 = \frac{[2][3][12]}{[4][5][6]}, \nn \\
 a_3 = -\frac{[2][3]}{[5][6]}, \ \ \ \
 & b_3=\frac{[2][3]}{[6]}=\frac{q^2+1+q^{-2}}{q^4+1+q^{-4}}=\frac{[3]}{[3]_2}, \ \ \
 & c_3=-\frac{[3]^2}{[5]}.
\ee

Similarly one look for the coefficients at higher representations using the ansatz

\be
C_{i_1\ldots i_r} = \frac{1}{[r+1]^2} \left( [2r+1]^{i_1}[2r-1]^{i_2}[2r-3]^{i_3}\ldots [3]^{i_r}
\ +\
\sum_{k=1}^r \  [2k+1]\cdot a_{k1}^{i_1}\cdot a_{k2}^{i_2} \cdot\ldots\cdot a_{kr}^{i_r}\right).
\ee
Again, this naively looks like a higher order system of equations, but it turns out possible to reduce it to at most quadratic. What helps is that the coefficients $a_{k1}$ have a relatively simple structure and their general form can be guessed:
\begin{eqnarray}
a_{11} & = & -\frac{[r]\cdot [2r+1]}{[r+2]}\,, \nn \\
a_{21} & = & \frac{[r-1]\cdot [r]\cdot [2r+1]}{[r+2]\cdot [r+3]}\,,  \nn \\
a_{31} & = & -\frac{[r-2][r-1][r]\cdot [2r+1]}{[r+2][r+3][r+4]}\,, \nn \\
&\cdots & \nn \\
a_{k1} & = & (-)^k \cdot \frac{[r]![r+1]!\cdot [2r+1]}{[r-k]! [r+k+1]!}\,.
\end{eqnarray}

Let us report the results that have been found for $r=4,5$ in the course of studying examples with genus $g$ up to $5$:

\paragraph{representation $[4]$:}
\begin{eqnarray}
a_{21} & = & [7]\cdot\frac{\{q\}^2}{q^2-1+q^{-2}}, \nn \\
a_{22} & = & -[2]^2\cdot \frac{q^4-q^2+1-q^{-2}+q^{-4}}{q^2-1+q^{-2}}, \nn \\
a_{23} & = & \frac{q^6+q^4+q^2-1+q^{-2}+q^{-4}+q^{-6}}{(q^4+q^{-4})(q^2-1+q^{-2})},\nn \\
a_{24} & = & -\frac{1}{(q^4+q^{-4})(q^2-1+q^{-2})}
\end{eqnarray}

\begin{eqnarray}
a_{31} & = & [5]\cdot\frac{q^4-q^2+1-q^{-2}+q^{-4}}{(q^2+q^{-2})(q^2-1+q^{-2})},\nn \\
a_{32} & = & \frac{[5]}{[7]}\cdot\frac{q^{10}-q^8-q^4-q^2+1-q^{-2}-q^{-4}-q^{-8}+q^{-10}}
{(q^2+q^{-2})(q^2-1+q^{-2})},\nn \\
a_{33} & = & -\frac{[5]}{[7]}\cdot\frac{(q^2+q^{-2})^2(q^4-q^2+1-q^{-2}+q^{-4})}{q^2-1+q^{-2}},\nn \\
a_{34} & = & \frac{[5]}{[7]}\cdot\frac{(q^2+q^{-2}) }{q^2-1+q^{-2}}
\end{eqnarray}

\begin{eqnarray}
a_{41} & = & \frac{q^6+2q^2-1+2q^{-2}+q^{-4}}{(q^2+q^{-2})(q^2-1+q^{-2})},\nn \\
a_{42} & = & \frac{q^6+q^2-1+q^{-2}+q^{-4}}{(q^2+q^{-2})(q^2-1+q^{-2})}
 =  [3]\cdot\frac{q^4-q^2+1-q^{-2}+q^{-4}}{(q^2+q^{-2})(q^2-1+q^{-2})} ,\nn \\
a_{43} & = & \frac{q^4-q^2-q^{-2}+q^{-4}}{q^2-1+q^{-2}}
= [3]\cdot\frac{\{q\}^2}{q^2-1+q^{-2}},\nn \\
a_{44} & = & -\frac{q^2+q^{-2}}{q^2-1+q^{-2}}.
\end{eqnarray}

\paragraph{representation $[5]$:}

\begin{eqnarray}
a_{21}&=&\frac{[9]([11]-[7]-[5]-[3]-1)}{[5][7]}\,, \nn
\\ a_{22} & = & -\frac{\left(q^2-q+1\right) \left(q^2+q+1\right) \left(q^6-q^3+1\right) \left(q^6+q^3+1\right) \left(q^{16}+2 q^{14}+q^{12}-q^8+q^4+2 q^2+1\right)}{q^6 \left(q^6-q^5+q^4-q^3+q^2-q+1\right) \left(q^6+q^5+q^4+q^3+q^2+q+1\right) \left(q^8+1\right)}\,, \nn
\\ a_{23} & = & \frac{[3]([11]+[9]+[7]-[5]-[3])}{[7]([5]-[3])}\,, \nn
\\  a_{24} & = &  \frac{-q^4 \left(q^2-q+1\right) \left(q^2+q+1\right) \left(q^{16}+q^{14}+q^{12}+q^{10}-q^8+q^6+q^4+q^2+1\right)}{\left(q^6-q^5+q^4-q^3+q^2-q+1\right) \left(q^6+q^5+q^4+q^3+q^2+q+1\right) \left(q^8+1\right) \left(q^8-q^6+q^4-q^2+1\right)}\,, \nn
\\ a_{25} & = & \frac{[3]}{[7]([5]-[3])([5]-2[3]+2)}\,.
\end{eqnarray}
\begin{eqnarray}
a_{31}&=&\frac{([11]+[9]-[5]-[3]-1)}{[5]}\,, \nn
\\ a_{32} & = & \frac{q^{28}+q^{26}-q^{24}-4 q^{22}-5 q^{20}-4 q^{18}-2 q^{16}-q^{14}-2 q^{12}-4 q^{10}-5 q^8-4 q^6-q^4+q^2+1}{q^6 \left(q^4-q^3+q^2-q+1\right) \left(q^4+q^3+q^2+q+1\right) \left(q^8+1\right)}\,, \nn
\\ a_{34} & = & \frac{([11]+[9]+[7]-[5]-[3])}{[11]-[9]+[5]-1}\,,\nn
\\ a_{35} & = & -\frac{([11]-[9]+[5]-1)}{[5]} \,.
\end{eqnarray}
\begin{eqnarray}
a_{41}&=&\frac{([11]+[9]+[7]-[3]-1)}{[7]}\,,\nn
\\ a_{42} & = & \frac{q^{28}+2 q^{26}+2 q^{24}-2 q^{20}-2 q^{18}+q^{16}+3 q^{14}+q^{12}-2 q^{10}-2 q^8+2 q^4+2 q^2+1}{q^4 \left(q^6-q^5+q^4-q^3+q^2-q+1\right) \left(q^6+q^5+q^4+q^3+q^2+q+1\right) \left(q^8+1\right)}\,,\nn
\\ a_{43} & = &-\frac{-q^{28}-q^{26}+q^{24}+4 q^{22}+5 q^{20}+4 q^{18}+2 q^{16}+q^{14}+2 q^{12}+4 q^{10}+5 q^8+4 q^6+q^4-q^2-1}{q^4 \left(q^{20}+q^{18}+q^{16}+q^{14}+2 q^{12}+2 q^{10}+2 q^8+q^6+q^4+q^2+1\right)}\,,\nn
\\ a_{44} & = & -\frac{[5]([9]+[7]-[5]-[3]-1)}{([5]-[3])[7]}\,,\nn
\\ a_{45} & = & \frac{[5]^2}{([5]-[3])[7]} \,.
\end{eqnarray}
\begin{eqnarray}
a_{51}&=&\frac{[3]([11]+[9]+[7]+[5]-1)}{[5][7]}\,,\nn
\\ a_{52} & = & \frac{[3]([11]+[9]+[7]-[3]-1)}{[5][7]}\,,\nn
\\ a_{53} & = & \frac{[3]([11]+[9]-[5]-[3]-1)}{[5][7]}\,,\nn
\\ a_{54} & = & \frac{[3]([11]-[7]-[5]-[3]-1)}{[5][7]}\,,\nn
\\ a_{55} & = & \frac{[3](-[9]-[7]-[5]-[3]-1)}{[5][7]} = -\frac{[3][5]}{[7]}\,.
\end{eqnarray}

We stress that the above formulae were obtained from the direct calculation of the invariants by the $S$-matrix method. They were originally used to derive the general formulae~(\ref{Jgrfirst}) and~(\ref{Arel}). While the latter remain a conjecture, the above results provide the colored Jones polynomials for genera $g\leq 5$ and representations $r\leq 5$, supporting the conjecture.

\end{document}